\documentclass{article}

\PassOptionsToPackage{numbers, compress}{natbib}
\usepackage[preprint]{Styles/neurips_2026}

\usepackage[utf8]{inputenc}   
\usepackage[T1]{fontenc}      
\usepackage{hyperref}         
\usepackage{url}              
\usepackage{booktabs}         
\usepackage{amsmath}          
\usepackage{amsfonts}         
\usepackage{amssymb}          
\usepackage{nicefrac}         
\usepackage{microtype}        
\usepackage{xcolor}           
\usepackage{graphicx}         
\usepackage{multirow}         
\usepackage{wrapfig}
\usepackage{float}

\title{SAC: Disaggregated KV Cache System for \\ Sparse Attention LLMs with CXL}

\author{%
  Ruiyang Ma$^{1,\ast}$\char44\ Teng Ma$^{2}$\char44\ Junru Li$^{2}$\char44\ Hantian Zha$^{3,\ast}$\char44\ Xuchun Shang$^{2}$\char44\ Qingda Hu$^{2}$\char44\\
  \ \textbf{Zheng Liu}$^{2}$\char44\ \textbf{Xinjun Yang}$^{2}$\char44\ \textbf{Tao Ma}$^{2}$\char44\ \textbf{Guojie Luo}$^{1}$ \\
  {
  $^{1}$Peking University \quad
  $^{2}$Alibaba Cloud \quad
  $^{3}$Renmin University of China}
}

\begin{document}

\maketitle

\begingroup
\renewcommand{\thefootnote}{*}
\footnotetext{This work was done while the author was an intern at Alibaba Cloud.}
\endgroup


\begin{abstract}

The scaling of LLMs toward long-context inference has shifted the primary serving system bottleneck from computation to memory capacity. 
Traditional solutions for dense attention models rely on RDMA-based disaggregated memory pools, which perform coarse-grained fetching of the entire prefix KV cache from remote storage to local memory before decoding.
However, this approach is fundamentally inefficient for emerging sparse attention models. While only a small fraction of KV entries are active during decoding, these systems still fetch the full KV cache locally, leading to severe transmission bottlenecks and local memory wastage.
To address this, we propose SAC, the first efficient disaggregated KV cache system optimized for sparse attention models. 
By leveraging the low-latency, cache-line granularity load/store semantics of Compute Express Link (CXL), SAC fetches only the required top-$k$ KV entries on demand during inference. 
Evaluations on DeepSeek-V3.2 using SGLang show that SAC achieves $2.1\times$ higher throughput, $9.7\times$ lower TTFT, and $1.8\times$ lower TBT compared to RDMA-based baselines, establishing CXL-based disaggregation as the superior infrastructure for emerging sparse attention models.

\end{abstract}

\section{Introduction}

The rapid scaling of LLM parameters and the growing demand for long-context inference shift the bottleneck from computation to memory, where prefix KV cache management has become a critical challenge in LLM serving~\cite{zhou2024survey}.
As local GPU HBM and DRAM fall short of the TB-level requirements of KV cache capacity, RDMA-based disaggregated KV cache systems emerge as the standard solution for memory expansion and cross-node sharing~\citep{mooncake, lmcache, NVIDIA2025Dynamo}.

In RDMA-based disaggregated systems, RDMA NICs prefetch the entire prefix KV cache into local memory to ensure low-latency access during the subsequent decoding stages of the request. Although this full fetching strategy is optimal for \textit{dense attention} models, it becomes fundamentally inefficient for the emerging generation of \textit{sparse attention} models, such as DeepSeek-V3.2~\citep{deepseekv32}, GLM-5.1~\citep{glm51} and DeepSeek-V4~\cite{deepseek_v4}.
The core problem is that only a small fraction of KV entries are active at any decoding step, while the KV cache for the full context must be fetched and remain in local memory during the request.
We find that the KV cache full fetching approach directly impacts the efficiency of disaggregated serving systems for sparse attention models due to two primary problems. 

\textbf{(P1) Transmission Bottleneck.}
Due to the significantly reduced computational complexity, the serving throughput of sparse attention models is typically bounded by the achievable batch size, rather than the compute capacity~\cite{zheng2026hisparse}. 
For long-context requests, KV cache size can reach dozens of GBs~\cite{sheng2023flexgen}. 
Maintaining high concurrency in sparse attention serving places substantial pressure on KV cache transmission, including RDMA transfers and memory layout rearrangement, which can cause long queuing delays and degrade time to first token and overall throughput.

\textbf{(P2) Local Memory Wasting.}
In sparse attention architectures, only the top-$k$ KV entries are utilized per layer for attention calculation. 
We observe that during the decoding of a long-context request, only a negligible portion of the prefix KV cache is actually accessed.
However, the entire prefix KV cache must still be fetched into local memory. 
To prevent local memory capacity from limiting the request batch size, TB-level memory is required to maintain high concurrency in sparse attention model serving, resulting in high infrastructure costs and poor memory efficiency.

An intuitive approach to solve the issue is to transmit only the top-$k$ KV cache entries on demand, which would reduce transmission pressure and conserve local memory.
However, this strategy is infeasible for RDMA-based disaggregated systems. 
First, the top-$k$ indices are dynamically determined per layer at runtime based on the current query vector. The process of fetching the top-$k$ KV is highly latency-sensitive, and the inherent access latency of RDMA cannot meet the strict timing requirement~\cite{zhu2024nanoflow}.
Second, the sparse top-$k$ KV entries consist of discrete, small data segments. Fetching these data via RDMA requires dozens of independent requests or complex gather/scatter operations, which incur significant software stack overhead~\cite{zhong2024distserve}.

Fortunately, the emerging \textbf{Compute Express Link (CXL)} technology~\cite{das2024introduction} introduces new opportunities for disaggregated KV cache systems in sparse attention models.
CXL is built on PCIe physical layer with a streamlined protocol stack, offering significantly lower latency than RDMA.
Moreover, CXL supports load/store semantics at cache-line granularity with zero message protocol overhead, making it uniquely suited for reading sparse and fine-grained data. Leveraging these features, we explore the feasibility of using CXL-based disaggregated memory to overcome the limitations of RDMA in sparse attention model serving.

We present \textbf{SAC} (Sparse Attention on CXL), which stores the KV cache in disaggregated CXL memory accessible to multiple servers. 
Due to its near-DRAM access latency and support for fine-grained load/store semantics, SAC enables the real-time fetching of top-$k$ KV entries during sparse attention calculations without significantly compromising serving performance. 
By transferring only the required top-$k$ entries on demand, SAC eliminates the transmission bottleneck (P1). Furthermore, because the full KV cache resides in the disaggregated CXL pool at all times, local memory is freed from maintaining redundant data, thereby resolving the local memory wasting problem (P2).

We integrate SAC into SGLang and evaluate on DeepSeek-V3.2 end-to-end serving across context lengths from 16K to 128K tokens. Compared with RDMA pool, SAC improves the decoding instance performance by achieving 2.1$\times$ higher throughput, 9.7$\times$ lower time to first token (TTFT), and 1.8$\times$ lower time between token (TBT). Compared with non-disaggregated upper bounds, SAC only incurs a 9\% throughput degradation, demonstrating the architectural superiority of CXL.

In summary, we make the following contributions:

\begin{itemize} 
    \item \textbf{Bottleneck Analysis.}
    We provide the first systematic analysis of the limitations inherent in traditional RDMA-based KV cache pooling for sparse attention models.

    \item \textbf{SAC System Design.}
    We design and implement the first CXL-based disaggregated KV cache system optimized for the serving of sparse attention models.

    \item \textbf{End-to-end Evaluation.}
    We integrate SAC into SGLang and evaluate using DeepSeek-V3.2. Our results demonstrate SAC significantly outperforms RDMA-based baselines, positioning CXL-based disaggregation as superior infrastructure for emerging sparse attention models.
\end{itemize}

\section{Background}

\subsection{Sparse Attention in LLMs}


Sparse attention is a recent advancement in Transformer architectures that reduces quadratic computational complexity to near-linear by restricting token attention to a relevant subset of the sequence. Recent LLMs have transitioned from training-free sparsity strategies~\citep{tang2024quest, zhang2023h2o, zhao2024buzz, li2024snapkv, streamingllm, liu2024clusterkv} toward native sparse attention architectures, such as DeepSeek-V3.2~\cite{deepseekv32}, GLM-5.1~\cite{glm51}, and DeepSeek-V4~\cite{deepseek_v4}, demonstrating the robustness and practicality of sparse attention for large-scale deployment.
The core mechanism driving this shift is \textit{DeepSeek Sparse Attention} (DSA)~\citep{deepseekv32}.

\begin{wrapfigure}{r}{0.53\textwidth} 
  \centering
  \includegraphics[width=\linewidth]{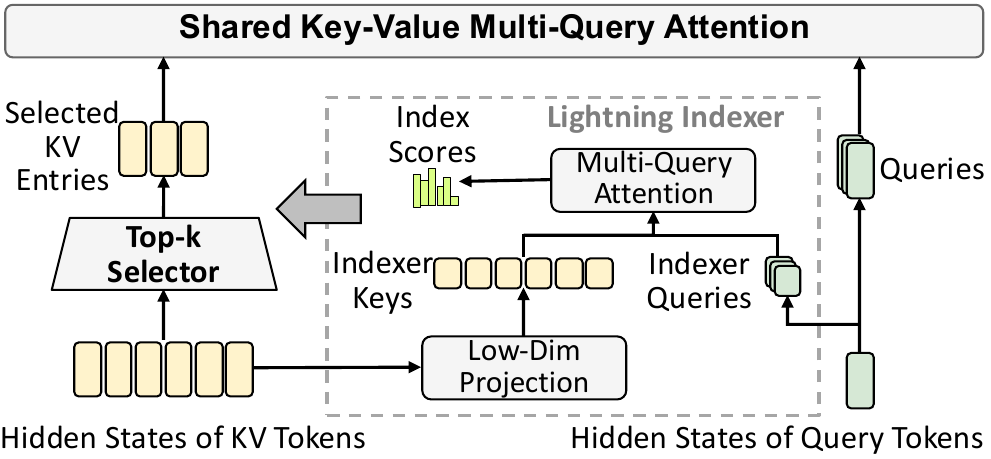}
  \vspace{-10pt}
  \caption{Workflow of DeepSeek Sparse Attention.}
  \vspace{-10pt}
  \label{fig:dsa}
\end{wrapfigure}

As illustrated in Figure~\ref{fig:dsa}, DSA employs a lightweight \textit{Lightning Indexer} to dynamically compute relevance scores using low-dimensional projections of latent KV vectors and queries. These projected 
indexer keys are computed once for each historical KV entry and stored in GPU memory for low-latency access. 
Based on these scores, a fine-grained selection mechanism retrieves only the top-$k$ ($2048$ in the current DSA implementation) KV latent vectors for Multi-head Latent Attention (MLA) calculations. Consequently, the computational cost of attention is reduced from $O(L^2)$ to $O(kL)$, enabling models to handle significantly longer input sequences and achieve higher serving throughput without compromising accuracy.



\subsection{Disaggregated Memory in LLM Serving}

\begin{wrapfigure}{r}{0.53\textwidth} 
  \centering
  \includegraphics[width=\linewidth]{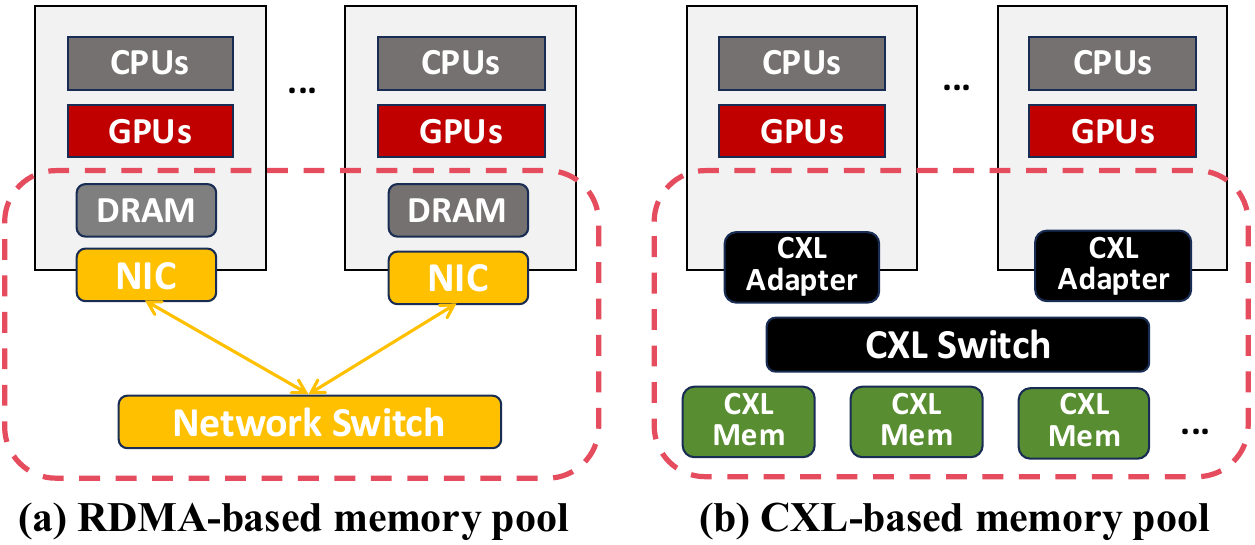}
  \vspace{-10pt}
  \caption{RDMA and CXL disaggregated memory.}
  \label{fig:rdma_cxl}
\end{wrapfigure}

The rapid scaling of LLM parameters and the escalating demand for long-context inference have made memory a critical bottleneck in GPU-accelerated serving systems, especially for the storage of KV cache. To address this, the industry is shifting toward RDMA-based disaggregated memory in LLM serving system to achieve memory capacity expansion and memory sharing across nodes~\cite{kvcache}, as shown in Figure~\ref{fig:rdma_cxl}(a).
Mainstream solutions, such as MoonCake~\cite{mooncake} and LMCache~\cite{lmcache}, use a CPU-driven RDMA access model to fetch remote data. The host CPU orchestrates the data transfers by moving the data from the GPU into an intermediate bounce buffer located in the host DRAM before finally issuing the RDMA request to send it to the remote pool.

While RDMA pools successfully extend memory capacity and facilitate prefix sharing across multi-server clusters, they exhibit inherent bottleneck when storing the KV cache for sparse attention models. 
RDMA transfers involve complex protocol stack including memory pinning, queue pair synchronization, context switching among others, leading to a high latency of several microseconds for each transfer~\cite{kalia2019datacenter}. Besides, the message-based send/recv semantics lead to significant performance degradation during small packet transfers~\citep{yang2025beluga}. 
For sparse KV access, where each layer needs small, independently chosen subset of KV entries, the overhead accumulates across all layers, making per-layer RDMA reads unacceptable for real-time top-k fetching for sparse KV cache.

\subsection{Compute Express Link (CXL)}

CXL is an emerging interconnect protocol built on the PCIe physical layer that facilitates high-bandwidth, low-latency memory sharing between hosts and accelerators~\cite{das2024introduction}. 
CXL supports memory expansion through Type-3 memory devices via the CXL.mem protocol. 
Modern CXL 2.0/3.0 switches~\cite{xconn_apollo_product} further enable multi-node memory sharing, providing a scalable hardware foundation for managing memory-intensive workloads.
As illustrated in Figure~\ref{fig:rdma_cxl}(b), CXL simplifies the data path compared to traditional RDMA by providing a hardware-managed load/store memory semantic. By enabling access at cache-line granularity, CXL is uniquely suited for fine-grained, sparse memory lookups, achieving remote memory access latencies that approach local DRAM.

Consequently, CXL has emerged as a compelling alternative to RDMA for memory pooling across diverse data center workloads, including databases~\cite{cxl_db_1, cxl_db_2, cxl_db_3} and cloud platforms~\cite{cxl_cloud_1, cxl_cloud_2, cxl_cloud_3}. Recent research also highlights its potential in LLM, particularly for KV cache management. 
Beluga~\citep{yang2025beluga} and TRACT~\citep{yoon2025tract} focus on prefix KV cache management for dense models with disaggregated CXL memory. In these systems, the entire prefix KV cache is prefetched into local memory like RDMA.
However, the KV cache access pattern in sparse attention models is fundamentally different, requiring per-layer reads of the sparse KV cache on demand. 
In the following section, we analyze the shortfall of existing RDMA-based disaggregated memory strategy and demonstrate the superiority of CXL.

\section{Motivation}

\vspace{-2.5pt}
\subsection{Bottlenecks of RDMA-based KV Cache System}
\vspace{-2.5pt}
\label{sec:bottlenecks}

Current RDMA-based disaggregated systems rely on KV cache full-fetching before decoding, which introduces two fundamental problems that hinder the serving efficiency of sparse attention models.

\begin{wrapfigure}{r}{0.32\textwidth} 
  \centering
  \includegraphics[width=\linewidth]{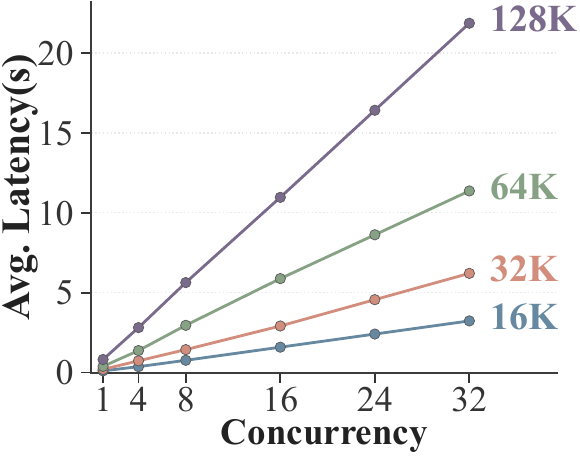}
  \vspace{-10pt}
  \caption{RDMA prefetch latency for prefix KV cache.}
  \label{fig:p1}
  \vspace{20pt}
  \centering
  \includegraphics[width=\linewidth]{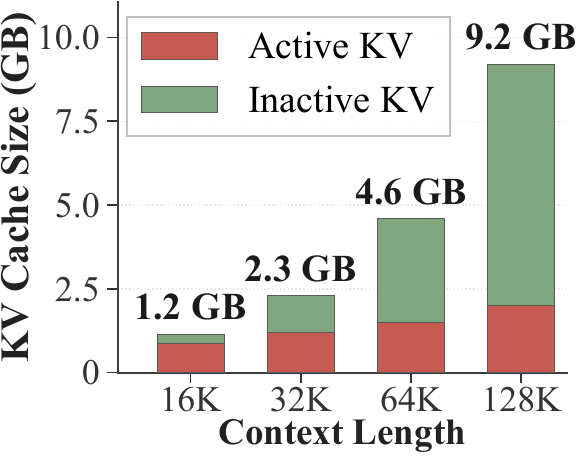}
  \vspace{-10pt}
  \caption{KV cache usage and footprint for DeepSeek-V3.2.}
  \vspace{-10pt}
  \label{fig:p2}
\end{wrapfigure}

\textbf{P1: Transmission Bottleneck.}
The transmission of massive KV cache blocks imposes significant pressure on RDMA bandwidth. Since a request cannot be processed until its entire prefix KV cache is retrieved, high-concurrency workloads frequently trigger network contention and substantial queuing delays. Additionally, KV caches are typically stored using a page-first layout on the remote but require a layer-first layout for local processing. This layout transformation introduces additional latency overhead.
Figure~\ref{fig:p1} illustrates the average latency for KV cache transmission under 100 Gbps RNIC. The latency grows near-linearly with concurrency and context length. In high-concurrency scenarios, the latency reach dozens of seconds, directly degrading the TTFT.
When transmission becomes the bottleneck, the system cannot retrieve enough requests for computation, limiting the serving throughput.

\textbf{P2: Local Memory Wasting.}
Fetching the full KV cache locally via RDMA is fundamentally inefficient for sparse attention models, as only a small fraction of the KV cache is active during decoding. Figure~\ref{fig:p2} profiles DeepSeek-V3.2 serving requests from the ShareGPT dataset~\cite{vicuna2023} with 1K-token output, the unused portion of the KV cache grows alongside context length. In 128K-context scenarios, only 21\% of the KV cache is actually utilized.
Moreover, the KV cache footprint for long context is substantial. DeepSeek-V3.2 requires 9.2 GB per request under a 128K context. Maintaining high concurrency for sparse attention model thus necessitates TB-level local memory. Consequently, full-fetching is not only unnecessary for sparse attention architectures but also imposes an excessive burden on local memory resources.

\begin{wrapfigure}{r}{0.55\textwidth} 
  \centering
  \includegraphics[width=\linewidth]{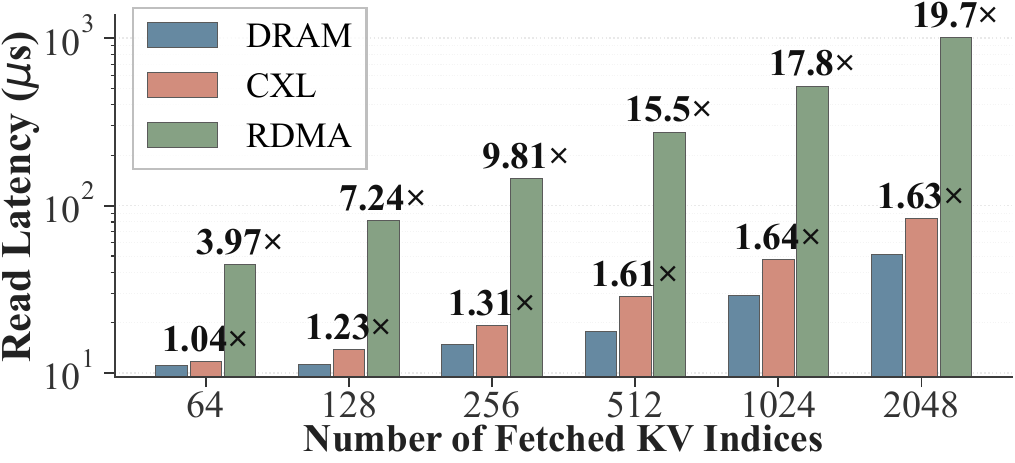}
  \vspace{-20pt}
  \caption{Retrieval latency comparison for sparse KV.}
  \vspace{-5pt}
  \label{fig:topk_profile}
\end{wrapfigure}

\vspace{-2.5pt}
\subsection{Memory Access Latency of Sparse Attention KV Cache}
\vspace{-2.5pt}
\label{sec:latency_profile}
To overcome the problems, sparse attention model serving requires high-efficiency, real-time loading of per-layer top-$k$ KV entries from disaggregated memory. In this section, we profile the retrieval latency for sparse KV data across different disaggregated memory architectures. Our experiment randomly samples sparse KV indices from a 128K-token context. Each entry corresponds to a DeepSeek-V3.2 MLA KV entry, which consists of a 512-dimensional latent KV vector and a 64-dimensional RoPE vector in \texttt{bfloat16} data type. We measure the GPU read latency across three memory backends: (i) disaggregated CXL memory, (ii) disaggregated RDMA memory, and (iii) local DRAM as baseline.

Figure~\ref{fig:topk_profile} illustrates the mean fetch latencies for varying numbers of sparse KV entries. 
CXL-based disaggregated memory achieves near-DRAM performance, maintaining a latency within $1.04$--$1.64\times$ of local DRAM. This demonstrates that CXL is an ideal fabric for sparse attention models. 
In contrast, RDMA exhibits significantly higher latency, ranging from $4.0$--$19.7\times$ slower than local DRAM and reaching millisecond-level delays as the number of KV entries increases. 
The results confirm that per-layer top-$k$ KV fetching via RDMA is impractical for inference, as the resulting latency would severely degrade performance. 
Consequently, in the following experiments, we exclude the RDMA dynamic top-$k$ fetch strategy from our comparisons. We focus on comparing the CXL-based strategy against the RDMA full-prefetch baseline to demonstrate the superiority of the CXL architecture.

\section{System Design}
\label{sec:design}

\subsection{SAC Workflow}
\label{sec:arch}
\begin{figure}[h] 
  \centering
  \includegraphics[width=\linewidth]{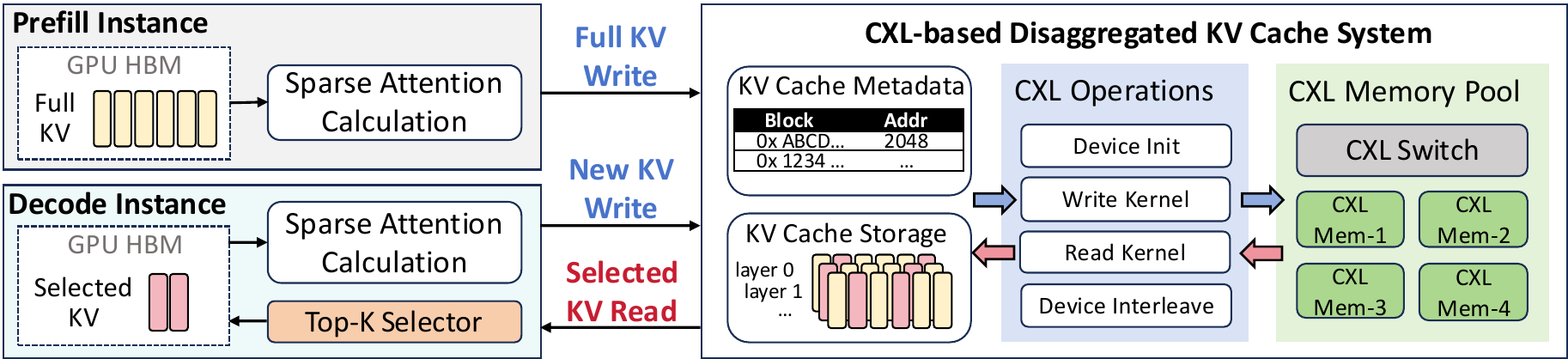}
  \vspace{-10pt}
  \caption{Overview of the SAC system and operational workflow, illustrating the interaction between compute instances and the disaggregated CXL memory pool.}
  \label{fig:sac}
\end{figure}

The SAC system transitions the KV cache management of sparse attention models from traditional RDMA-based architectures to a CXL-based disaggregated architecture, which is designed to efficiently meet the high-concurrency demands of disaggregated serving for sparse attention models. The system architecture comprises three primary components: the Prefill Instance, the Decode Instance, and the CXL-based Disaggregated KV Cache System, as shown in Figure~\ref{fig:sac}.

\textbf{Prefill Instance. } 
The prefill stage is characterized by large batch sizes and is primarily compute-bound. During prefilling, full KV caches are maintained in local GPU memory to ensure fast access. The prefill instance is responsible for populating the CXL memory pool with the calculated KV entries upon completion. Because the process of writing full KV blocks to CXL memory is well-established in prior work~\cite{yang2025beluga, yoon2025tract} and is not unique to sparse attention models, it is not the focus of this paper.

\textbf{Decode Instance. } 
The decode instance is built upon HiSparse framework in SGLang~\cite{zheng2026hisparse} (detailed in Appendix~\ref{app:hisparse}), which is designed for high-throughput inference for sparse attention models. The core mechanism involves offloading the KV cache to low-tiered memory and using swap-in kernels to dynamically load only the top-$k$ KV into GPU for attention calculations. SAC extends this framework by integrating disaggregated memory backend and CXL-based KV cache management system. 
During each attention layer, SAC directly fetches the required top-$k$ KV from CXL memory pool into GPU. The KV entries for newly generated tokens are written back to the CXL memory.

\textbf{CXL-based Disaggregated KV cache System. } 
The CXL-based system manages the KV cache and its associated metadata, mapping all data into a global CXL address space accessible by multiple compute instances. It handles CXL device initialization and provides optimized read/write kernels and device interleaving strategy.
The physical memory pool consists of multiple CXL devices interconnected via a CXL switch, which provides fine-grained, low-latency access required to support the sparse read and write patterns inherent in sparse attention inference.

\subsection{SAC System Topology}

\begin{wrapfigure}{r}{0.55\textwidth} 
  \centering
  \vspace{-10pt}
  \includegraphics[width=\linewidth]{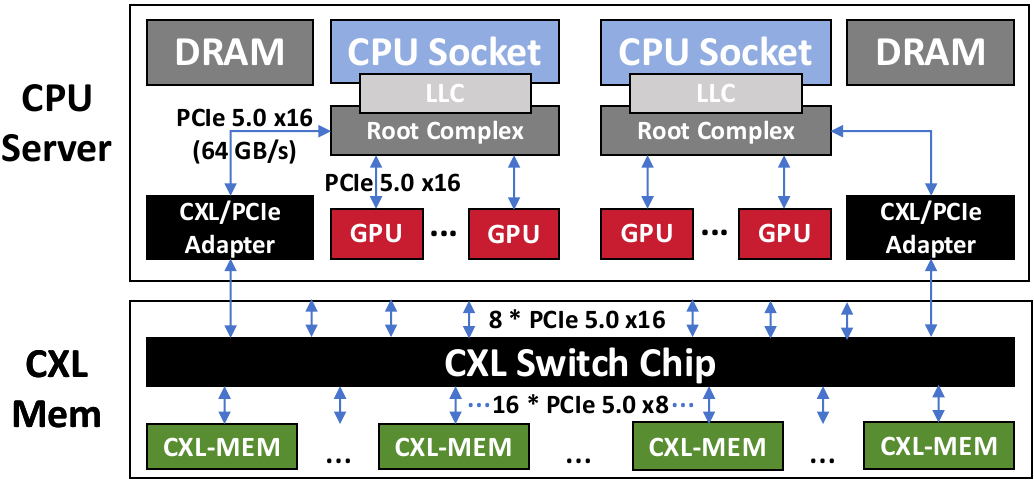}
  \vspace{-15pt}
  \caption{Hardware topology of SAC system.}
  \vspace{-10pt}
  \label{fig:cxl_arch}
\end{wrapfigure}

The SAC system relies on a CXL-based disaggregated architecture to provide the fine-grained, low-latency access required for sparse attention. 
Figure~\ref{fig:cxl_arch} illustrates the architecture overview of the memory system. Each server in the cluster contains host CPUs and multiple GPUs. Each NUMA node is connected to the CXL switch via a PCIe 5.0 x16 PCIe/CXL adapter.
The CXL memory pool consists of a switch node, which is equipped with XConn XC50256 CXL switch chip~\cite{xconn_apollo_product}. The chip has 256 PCIe 5.0 lanes which are partitioned between CXL memory devices and compute servers. The CXL switch chip provides 512 GB/s of total bandwidth between memory devices and compute servers, which connects up to 8 servers to the CXL memory pool. 
By performing hardware-level address mapping and forwarding, this architecture enables SAC to achieve robust multi-host scaling, allowing multiple GPUs to retrieve non-contiguous top-$k$ entries with near-DRAM latency while bypassing the protocol overhead inherent in RDMA-based message-passing architectures.

\subsection{CXL-based KV Cache Management}
\label{sec:cxl-impl}

\subsubsection{Unified CXL Memory Resource} 
\label{sec:interleave}
The CXL disaggregated memory serves as a unified resource for both the KV cache data and the metadata index. Compared to RDMA-based architectures, CXL approach offers two advantages.

\textbf{Locality-Transparent KV Cache Layout.}
Traditional RDMA-based disaggregated systems suffer from a significant performance gap between local and remote memory~\citep{mooncake, NVIDIA2025Dynamo}, necessitating complex and cache-aware scheduling policies. 
In contrast, SAC leverages CXL to achieve near-DRAM latency.
Since CXL supports byte-addressable, fine-grained load/store semantics, we adopt a locality-transparent, layer-first memory layout that is identical to the structure used in local GPU.
This allows GPU to access KV cache in CXL memory using the same logic as local memory. 
For sparse attention models, per-layer top-$k$ retrieval is transformed into lightweight memory read operations in the GPU kernels, bypassing complex buffer management and reducing scheduling complexity.

\textbf{CXL-based Metadata Management.}
Conventional shared KV cache systems typically rely on centralized metadata services accessed via high-overhead protocols such as RDMA or TCP/IP~\cite{yang2025beluga}.
SAC replaces these expensive network-based interactions by hosting metadata within a dedicated, globally accessible CXL shared-memory region.
By utilizing native load/store semantics for cross-server synchronization, SAC replaces heavy remote procedure call (RPC) interactions with high-speed memory operations. This approach ensures that metadata updates and lookups are performed with near-DRAM latency and eliminates kernel-mode transitions and context switches, providing a responsive control plane for high-concurrency sparse attention workloads.

\subsubsection{CXL Operation Implementation}

\textbf{CXL Device Initialization.}
CXL devices are exposed as DAX (Direct Access) devices in operating system. A \texttt{mmap} system call maps the DAX device to the host virtual address. By using \texttt{cudaHostRegister}, GPU accesses CXL address with the same load/store instructions for DRAM. 

\textbf{GPU Write Path.}
The prefill instance transfers the KV cache in GPU to CXL memory through a layer-wise offloading strategy, which is executed by a memory-coalesced kernel where each warp is assigned a designated subset of KV indices. By using vectorized store instructions (\texttt{st.global.b64}), the kernel achieves high-bandwidth, single-pass writes into the CXL global address space.

\textbf{GPU Read Path.}
The decoding instance retrieves the top-$k$ KV required for each attention layer from the CXL memory. The read path is optimized for sparse, non-contiguous access to multiple fixed-length vectors. With a coalesced kernel, SAC uses vectorized load instructions (\texttt{ld.global.b64}) to efficiently fetch the data into GPU, satisfying the real-time constraints of sparse attention inference.

\subsubsection{CXL Bandwidth Optimization}
\label{sec:cxl_interleave}
\begin{wrapfigure}{r}{0.2\textwidth} 
  \centering
  \includegraphics[width=\linewidth]{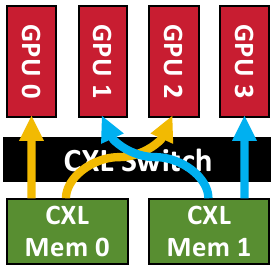}
  \caption{CXL interleaving example.}
  \vspace{-10pt}
  \label{fig:cxl_interleave}
\end{wrapfigure}
The CXL memory pool is composed of multiple devices, each interconnected via a PCIe 5.0 x8 link to the switch. To optimize upstream bandwidth, it is necessary to balance data traffic across these devices. For sparse attention models based on MLA, data parallel attention is a recommended configuration to avoid replicated KV cache storage. In this setup, each request's attention is calculated on only one GPU. The challenge is to reduce the contention caused by multiple GPUs concurrently accessing the same CXL device.

In SAC, we store the KV cache of a single request within one CXL device. When dispatching requests to model runners, the scheduler considers their distribution across the CXL pool to ensure GPUs access different devices in an interleaved manner. We assign GPU rank to CXL devices in a round-robin fashion, as shown in Figure~\ref{fig:cxl_interleave}. By distributing traffic across multiple physical links during parallel GPU access, this approach alleviates single-link pressure and significantly reduces retrieval latency.

\section{Evaluation}
\label{sec:eval}

We integrate SAC into SGLang~\cite{zheng2024sglang} and HiSparse framework~\cite{zheng2026hisparse}. 
We evaluate the DeepSeek-V3.2 model~\cite{deepseekv32} (AWQ 4-bit quantization) using an 8-H20 GPU server and 2TB CXL memory pool as KV cache backend. 
We compare SAC against two alternative backends: RDMA-based memory pool and local host DRAM as upper-bound baseline.
Detailed experimental setup is provided in Appendix~\ref{app:exp-details}. 

\subsection{End-to-End Performance Comparison} 
\label{sec:e2e}
We evaluate the end-to-end performance of SAC across three memory backends. To isolate the performance impact on the prefill and decoding stages, we design a two-round experimental setup:
\begin{itemize}
\item \textbf{Round-1 (Cache Populate):} This scenario evaluates the prefill instance. The initial KV cache is computed on the GPU and subsequently stored in the disaggregated memory pool.
\item \textbf{Round-2 (Cache Hit)}: This scenario evaluates the decode instance, where all the KV cache is pre-populated in the memory pool and the prefill stage is bypassed entirely.
\end{itemize}
We measure serving metrics including output token throughput, time between tokens (TBT) and time to first token (TTFT). The benchmark uses 512 requests sampled from the ShareGPT dataset~\cite{vicuna2023}, with concurrency set to 8 for Round-1 and 64 for Round-2. 
We sweep the context length from 16K to 128K tokens and fix the output length at 1K tokens. Additional results with varying output lengths are provided in Appendix~\ref{app:output_len}.

\begin{figure}[h]
\centering
\includegraphics[width=0.99\linewidth]{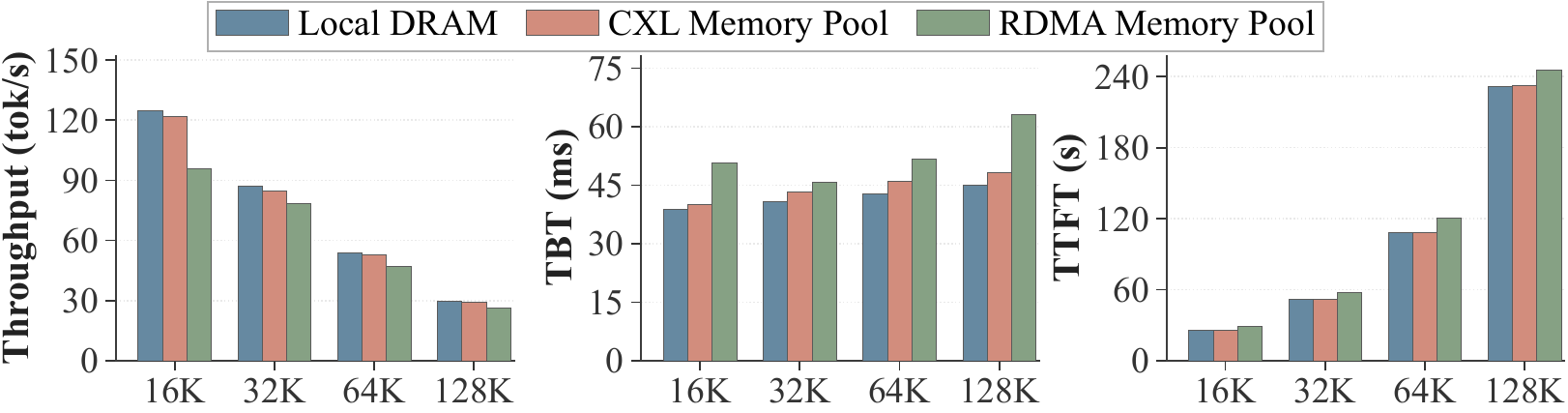} 
\vspace{-0.5em} 
\caption{Round-1 performance comparison of SAC against RDMA and local DRAM backend.}
\label{fig:exp_main_1}
\end{figure}

\begin{figure}[h]
\centering
\includegraphics[width=0.99\linewidth]{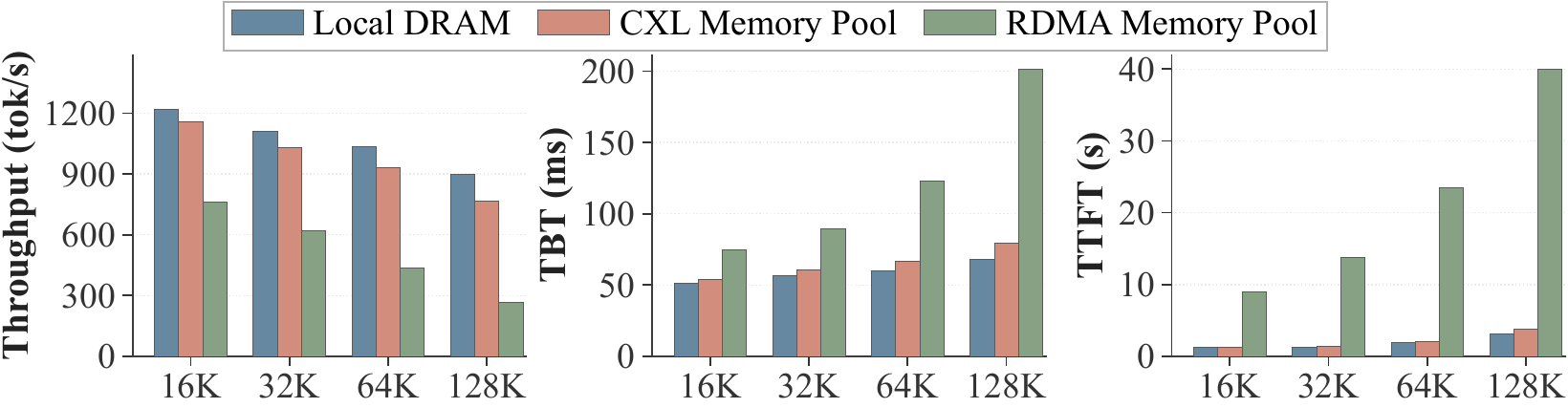} 
\vspace{-0.5em} 
\caption{Round-2 performance comparison of SAC against RDMA and local DRAM backend.}
\label{fig:exp_main_2}
\end{figure}

As illustrated in Figure~\ref{fig:exp_main_1} and~\ref{fig:exp_main_2}, SAC closely approximates the performance of the local DRAM upper bound in both Round-1 and Round-2. 
On average, SAC achieves a throughput within 91\% of the DRAM baseline, with only a marginal increase in TBT and TTFT. 
In Round-1 prefill stage, since both CXL and RDMA maintain the full KV cache on the GPU for computation and subsequently store the KV cache into the memory pool, the two backends exhibit comparable performance.
In Round-2, SAC significantly outperforms RDMA baseline, delivering on average $2.1\times$ higher throughput. RDMA suffers from a substantially higher $9.7\times$ TTFT than SAC, as high client concurrency saturates the transmission bandwidth during KV retrieval, resulting in increased queuing delays. Besides, the TBT of RDMA is $1.8\times$ higher than SAC due to severe PCIe bus contention, which must simultaneously handle incoming KV cache traffic and the HiSparse swap-in process.


\subsection{Throughput Scalability Comparison} 
\label{sec:transmission}

We evaluate the decoding throughput across varying concurrency to demonstrate the scalability of SAC and the transmission bottleneck inherent in RDMA-based systems.  
We compare SAC with RDMA backend at context lengths of 32K, 64K, and 128K.

\begin{figure}[h]
\centering
\includegraphics[width=\linewidth]{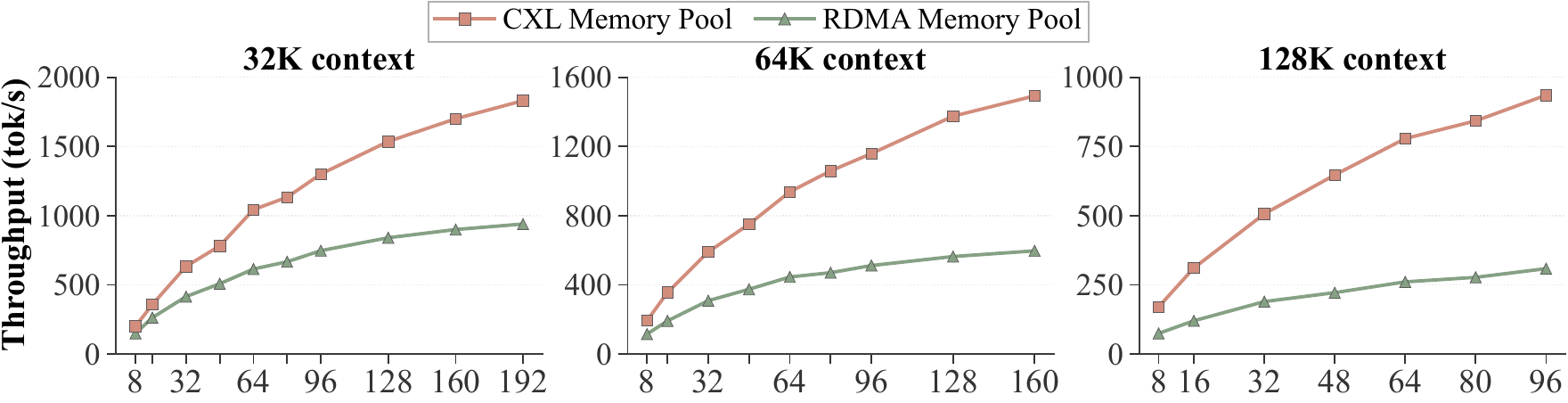} 
\vspace{-0.5em} 
\caption{Decoding throughput scalability comparison of SAC and RDMA under concurrency.}
\label{fig:exp_scalability}
\end{figure}

As illustrated in Figure~\ref{fig:exp_scalability}, SAC exhibits robust throughput scalability, with performance increasing consistently alongside concurrency. For 32K, 64K, and 128K, SAC achieves up to $2.0\times$, $2.5\times$ and $3.1\times$ higher throughput than the RDMA-based system.
In contrast, RDMA backend quickly reaches a throughput plateau due to the full fetching of KV cache, which exhausts the transmission bandwidth. 
SAC bypasses this bottleneck by leveraging CXL fine-grained load/store semantics to fetch only the necessary top-$k$ entries. This significantly reduces the per-request bandwidth footprint, allowing SAC to sustain high-throughput decoding even as context length and concurrency scale.

\subsection{Comparison with Non-Disaggregated Baselines} 
\label{sec:concurrency}
We compare the decoding throughput of SAC against non-disaggregated configurations to demonstrate the architectural superiority of CXL memory. We evaluate SAC against local DRAM and GPU-only baselines, where KV cache is stored in host memory and GPU HBM respectively.

\begin{figure}[h]
\centering
\includegraphics[width=\linewidth]{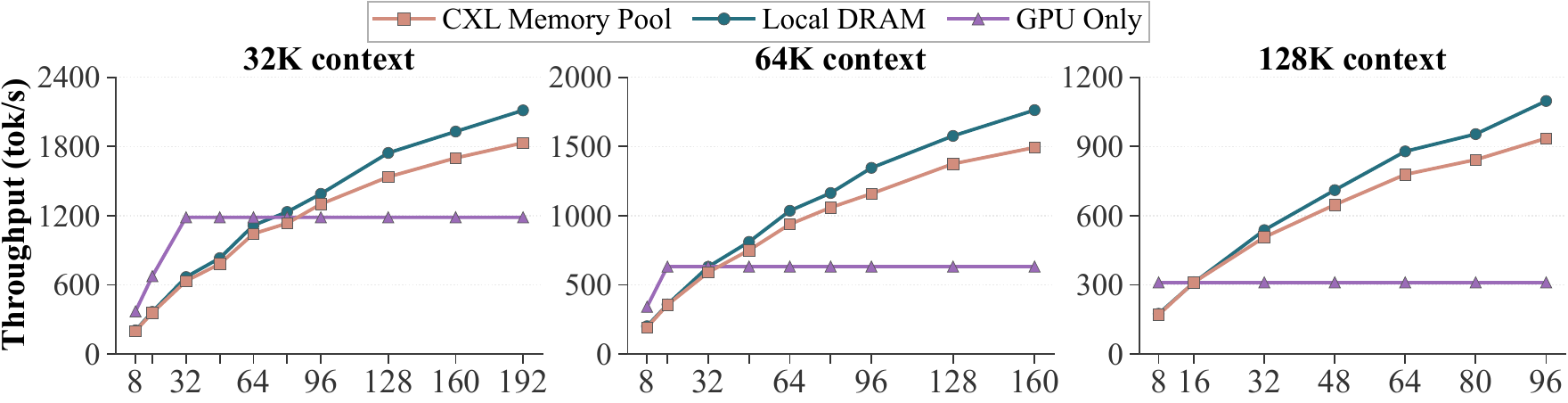} 
\vspace{-0.5em} 
\caption{Throughput comparison of SAC with non-disaggregated baselines under concurrency.}
\label{fig:exp_nodis}
\end{figure}

As illustrated in Figure~\ref{fig:exp_nodis}, although SAC stores the KV cache in disaggregated CXL memory, it achieves performance close to the local DRAM baseline.
When compared to the HBM-only baseline, the GPU HBM delivers higher throughput at low concurrency. However, as concurrency increases, the HBM reaches a capacity limit where the maximum achievable batch size cannot increase further. This demonstrates the importance of using lower-tier memory for sparse attention inference, which provides the necessary capacity to scale to high-concurrency workloads.

\subsection{Impact of CXL Device Interleaving}
\label{sec:interleave-exp}

\begin{wrapfigure}{r}{0.32\textwidth} 
  \centering
  \includegraphics[width=\linewidth]{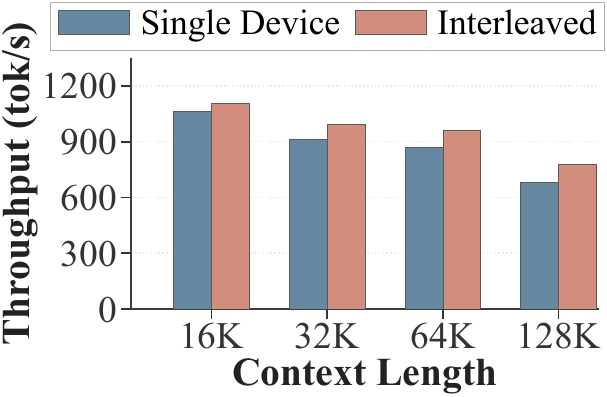}
  \vspace{-15pt}
  \caption{Throughput gain from CXL device interleaving.}
  \label{fig:exp_interleave}

  \centering
  \vspace{+20pt}
  \includegraphics[width=\linewidth]{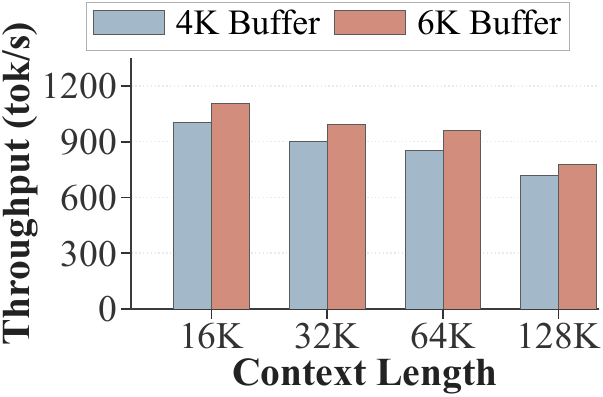}
  \vspace{-15pt}
  \caption{Impact of GPU device buffer size.}
  \vspace{-20pt}
  \label{fig:exp_buffer}
\end{wrapfigure}

We conduct an ablation study to evaluate the effectiveness of device-aware interleaving strategy introduced in Section~\ref{sec:cxl_interleave}. We compare the decoding throughput using a single CXL memory device against an interleaved configuration across two CXL devices. 
As illustrated in Figure~\ref{fig:exp_interleave}, the interleaved configuration consistently outperforms the single-device baseline. On average, interleaving improves decoding throughput by 9.2\%, with a peak gain of 14.2\% at 128K context 
length. This demonstrates that by interleaving KV cache storage across multiple devices, SAC effectively mitigates link-level contention. The results suggest that scaling to additional CXL devices would further bridge the performance gap between SAC and the local DRAM baseline by providing greater aggregate bandwidth to sustain high-performance inference for long-context workloads.

\subsection{Impact of HiSparse Configuration}
SAC is built upon HiSparse, where a critical configuration parameter is \texttt{device\_buffer\_size}. This parameter determines the capacity of the hot KV cache stored in GPU HBM, which directly impacts the volume of KV cache data transmitted from disaggregated memory to the GPU. We evaluate two configurations: \texttt{device\_buffer\_size} set to 4K and 6K.


As illustrated in Figure~\ref{fig:exp_buffer}, SAC performance improves as the buffer size increases; the 6K configuration achieves an average throughput 10.4\% higher than the 4K baseline. This improvement is driven by a lower KV cache miss rate, which reduces both the total transmission data size and the resulting pressure on the CXL link. These results suggest that maintaining a larger GPU buffer is beneficial for SAC performance, as it minimizes the overhead of data transfers from the disaggregated memory.

\section{Discussion}
\label{sec:discussion}
\textbf{Generality to DeepSeek-V4.}
The latest DeepSeek-V4~\cite{deepseek_v4} hybridly uses compressed sparse attention and heavily compressed attention to support context lengths up to 1M tokens. Since its sparse attention shares the same top-$k$ access pattern as DeepSeek-V3.2, it can directly benefit from SAC. Compared to RDMA, CXL supports more flexible storage and access patterns for heterogeneous KV caches through load/store semantics, showing great potential as a backend for such models.

\textbf{Memory Pool Paradigm Transform.}
The evolution of model sparsity is driving KV cache access toward a fine-grained, heterogeneous pattern.
Traditional message-based protocols cannot satisfy the requirements.
LLM serving clusters are evolving toward memory-semantic interconnects, necessitating tighter coupling between compute and memory nodes to maximize performance.
CXL serves as a cornerstone for this transition and the emergence of unified scale-up protocols \cite{liao2025ub, arsid2025ultra} promises more robust solutions for high-performance, disaggregated KV cache storage.

\textbf{Limitation and Future Work}
This paper focuses on experimenting with DeepSeek-V3.2 as the most representative sparse attention model. Evaluating additional models, such as GLM-5.1~\cite{glm51} and DeepSeek-V4~\cite{deepseek_v4}, remains a task for future work. Additionally, further optimizations can be applied to SAC. For instance, better utilizing the memory hierarchy, including HBM and DRAM, to organize the KV cache is a promising direction for improving system efficiency.

\section{Conclusion}

In this paper, we address the fundamental inefficiency of traditional RDMA-based KV cache system when serving emerging sparse attention models. 
We identify that the KV cache full-fetching in RDMA-based systems leads to transmission bottlenecks and local memory waste. 
To overcome the limitations, we present SAC, a disaggregated KV cache system that leverages the low-latency, fine-grained load/store semantics of CXL.
By fetching only the required top-$k$ KV entries on demand, SAC effectively eliminates the transmission overhead and enables high-concurrency serving without the burden of TB-level local memory. 
Our end-to-end evaluation with DeepSeek-V3.2 demonstrates that SAC achieves 2.1$\times$ higher throughput, 9.7$\times$ lower TTFT, and 1.8$\times$ lower TBT compared to RDMA-based baselines. 
As LLM architectures continue to evolve toward higher sparsity and longer contexts, SAC demonstrates that CXL-based disaggregation provides the superior architectural foundation necessary for the next generation of scalable and efficient LLM serving infrastructure.


\newpage
\bibliographystyle{unsrtnat}
\bibliography{references}

@article{deepseekv32,
  title={Deepseek-v3. 2: Pushing the frontier of open large language models},
  author={Liu, Aixin and Mei, Aoxue and Lin, Bangcai and Xue, Bing and Wang, Bingxuan and Xu, Bingzheng and Wu, Bochao and Zhang, Bowei and Lin, Chaofan and Dong, Chen and others},
  journal={arXiv preprint arXiv:2512.02556},
  year={2025}
}

@misc{glm51,
  author = {Zhipu AI},
  title = {GLM-5.1},
  year = {2026},
  publisher = {Hugging Face},
  journal = {Hugging Face Repository},
  howpublished = {\url{https://huggingface.co/zai-org/GLM-5.1}},
}

@misc{deepseek_v4,
  author = {{DeepSeek-AI}},
  title = {DeepSeek-V4-Pro},
  year = {2026},
  publisher = {Hugging Face},
  journal = {Hugging Face Repository},
  howpublished = {\url{https://huggingface.co/deepseek-ai/DeepSeek-V4-Pro}},
  note = {Accessed: 2026-05-02}
}

@article{zhang2023h2o,
  title={H2o: Heavy-hitter oracle for efficient generative inference of large language models},
  author={Zhang, Zhenyu and Sheng, Ying and Zhou, Tianyi and Chen, Tianlong and Zheng, Lianmin and Cai, Ruisi and Song, Zhao and Tian, Yuandong and R{\'e}, Christopher and Barrett, Clark and others},
  journal={Advances in Neural Information Processing Systems},
  volume={36},
  pages={34661--34710},
  year={2023}
}

@article{streamingllm,
  title={Efficient streaming language models with attention sinks},
  author={Xiao, Guangxuan and Tian, Yuandong and Chen, Beidi and Han, Song and Lewis, Mike},
  journal={arXiv preprint arXiv:2309.17453},
  year={2023}
}

@article{zhao2024buzz,
  title={Buzz: Beehive-structured sparse kv cache with segmented heavy hitters for efficient {LLM} inference},
  author={Zhao, Junqi and Fang, Zhijin and Li, Shu and Yang, Shaohui and He, Shichao},
  journal={arXiv preprint arXiv:2410.23079},
  year={2024}
}

@article{li2024snapkv,
  title={{SnapKV}: {LLM} knows what you are looking for before generation},
  author={Li, Yuhong and Huang, Yingbing and Yang, Bowen and Venkitesh, Bharat and Locatelli, Acyr and Ye, Hanchen and Cai, Tianle and Lewis, Patrick and Chen, Deming},
  journal={Advances in Neural Information Processing Systems},
  volume={37},
  pages={22947--22970},
  year={2024}
}

@inproceedings{tang2024quest,
  title={QUEST: query-aware sparsity for efficient long-context {LLM} inference},
  author={Tang, Jiaming and Zhao, Yilong and Zhu, Kan and Xiao, Guangxuan and Kasikci, Baris and Han, Song},
  booktitle={Proceedings of the 41st International Conference on Machine Learning},
  pages={47901--47911},
  year={2024}
}

@article{liu2024clusterkv,
  title={Clusterkv: Manipulating {LLM} {KV} cache in semantic space for recallable compression},
  author={Liu, Guangda and Li, Chengwei and Zhao, Jieru and Zhang, Chenqi and Guo, Minyi},
  journal={arXiv preprint arXiv:2412.03213},
  year={2024}
}

@article{zheng2024sglang,
  title={Sglang: Efficient execution of structured language model programs},
  author={Zheng, Lianmin and Yin, Liangsheng and Xie, Zhiqiang and Sun, Chuyue and Huang, Jeff and Yu, Cody H and Cao, Shiyi and Kozyrakis, Christos and Stoica, Ion and Gonzalez, Joseph E and others},
  journal={Advances in neural information processing systems},
  volume={37},
  pages={62557--62583},
  year={2024}
}

@misc{zheng2026hisparse,
  author = {Zhiqiang Xie, Zhangheng Huang, Tingwei Huang},
  title = {HiSparse: High-Efficiency Sparse Attention Inference in SGLang},
  year = {2026},
  month = {April},
  howpublished = {\url{https://www.lmsys.org/blog/2026-04-10-sglang-hisparse/}},
  note = {LMSYS Blog}
}

@article{kvcache,
  title={{KVDirect}: Distributed disaggregated {LLM} inference},
  author={Chen, Shiyang and Jiang, Rain and Yu, Dezhi and Xu, Jinlai and Chao, Mengyuan and Meng, Fanlong and Jiang, Chenyu and Xu, Wei and Liu, Hang},
  journal={arXiv preprint arXiv:2501.14743},
  year={2024}
}

@inproceedings{mooncake,
  title={Mooncake: Trading more storage for less computation a {KVCache}-centric architecture for serving {LLM} chatbot},
  author={Qin, Ruoyu and Li, Zheming and He, Weiran and Cui, Jialei and Ren, Feng and Zhang, Mingxing and Wu, Yongwei and Zheng, Weimin and Xu, Xinran},
  booktitle={23rd USENIX Conference on File and Storage Technologies (FAST 25)},
  pages={155--170},
  year={2025}
}

@inproceedings{lmcache,
  title={{CacheGen}: {KV} cache compression and streaming for fast large language model serving},
  author={Liu, Yuhan and Li, Hanchen and Cheng, Yihua and Ray, Siddhant and Huang, Yuyang and Zhang, Qizheng and Du, Kuntai and Yao, Jiayi and Lu, Shan and Ananthanarayanan, Ganesh and others},
  booktitle={Proceedings of the ACM SIGCOMM 2024 Conference},
  pages={38--56},
  year={2024}
}

@misc{NVIDIA2025Dynamo,
  author       = {{NVIDIA Corporation}},
  title        = {{NVIDIA Dynamo Open-Source Library Accelerates and Scales AI Reasoning Models}},
  howpublished = {\url{https://nvidianews.nvidia.com/news/nvidia-dynamoopen-source-library-accelerates-and-scales-ai-reasoningmodels}},
  year         = {2025}
}

@article{yang2025beluga,
  title={Beluga: A cxl-based memory architecture for scalable and efficient {LLM} kvcache management},
  author={Yang, Xinjun and Hu, Qingda and Li, Junru and Li, Feifei and Zhu, Yicong and Zhou, Yuqi and Lin, Qiuru and Dai, Jian and Kong, Yang and Zhang, Jiayu and others},
  journal={Proceedings of the ACM on Management of Data},
  volume={4},
  number={1 (SIGMOD},
  pages={1--29},
  year={2026},
  publisher={ACM New York, NY, USA}
}

@article{yoon2025tract,
  title={{TraCT}: Disaggregated {LLM} Serving with {CXL} Shared Memory {KV} Cache at Rack-Scale},
  author={Yoon, Dongha and Min, Younghoon and Kim, Hoshik and Noh, Sam H and Kim, Jongryool},
  journal={arXiv preprint arXiv:2512.18194},
  year={2025}
}

@article{das2024introduction,
  title={An introduction to the compute express link ({CXL}) interconnect},
  author={Das Sharma, Debendra and Blankenship, Robert and Berger, Daniel},
  journal={ACM Computing Surveys},
  volume={56},
  number={11},
  pages={1--37},
  year={2024},
  publisher={ACM New York, NY}
}

@misc{xconn_apollo_product,
  author       = {{XConn Technologies}},
  title        = {{XC50256}: World's First Hybrid {CXL} 2.0 and {PCIe} {Gen5} Switch {IC}},
  howpublished = {\url{https://www.xconn-tech.com/product}},
  year         = {2023},
}

@inproceedings{cxl_db_1,
  title={Unlocking the potential of {CXL} for disaggregated memory in cloud-native databases},
  author={Yang, Xinjun and Zhang, Yingqiang and Chen, Hao and Li, Feifei and Fan, Gerry and Kong, Yang and Wang, Bo and Fang, Jing and Wang, Yuhui and Huang, Tao and others},
  booktitle={Companion of the 2025 International Conference on Management of Data},
  pages={689--702},
  year={2025}
}

@inproceedings{cxl_db_2,
  title={Tigon: A Distributed Database for a {CXL} Pod},
  author={Huang, Yibo and Chen, Haowei and Ni, Newton and Sun, Yan and Chidambaram, Vijay and Tang, Dixin and Witchel, Emmett},
  booktitle={19th USENIX Symposium on Operating Systems Design and Implementation (OSDI 25)},
  pages={109--128},
  year={2025}
}

@article{cxl_db_3,
  title={Rcmp: Reconstructing {RDMA}-based memory disaggregation via {CXL}},
  author={Wang, Zhonghua and Guo, Yixing and Lu, Kai and Wan, Jiguang and Wang, Daohui and Yao, Ting and Wu, Huatao},
  journal={ACM Transactions on Architecture and Code Optimization},
  volume={21},
  number={1},
  pages={1--26},
  year={2024},
  publisher={ACM New York, NY}
}

@inproceedings{cxl_cloud_1,
  title={Managing memory tiers with {CXL} in virtualized environments},
  author={Zhong, Yuhong and Berger, Daniel S and Waldspurger, Carl and Wee, Ryan and Agarwal, Ishwar and Agarwal, Rajat and Hady, Frank and Kumar, Karthik and Hill, Mark D and Chowdhury, Mosharaf and others},
  booktitle={18th USENIX Symposium on Operating Systems Design and Implementation (OSDI 24)},
  pages={37--56},
  year={2024}
}

@article{cxl_cloud_2,
  title={Memtunnel: A {CXL}-based rack-scale host memory pooling architecture for cloud service},
  author={Guan, Tianchan and Guan, Yijin and Du, Zhaoyang and Ma, Jiacheng and Tian, Boyu and Wang, Zhao and Ma, Teng and Liu, Zheng and Kong, Yang and Xie, Yuan and others},
  journal={IEEE Transactions on Parallel and Distributed Systems},
  year={2025},
  publisher={IEEE}
}

@inproceedings{cxl_cloud_3,
  title={Pond: {CXL}-based memory pooling systems for cloud platforms},
  author={Li, Huaicheng and Berger, Daniel S and Hsu, Lisa and Ernst, Daniel and Zardoshti, Pantea and Novakovic, Stanko and Shah, Monish and Rajadnya, Samir and Lee, Scott and Agarwal, Ishwar and others},
  booktitle={Proceedings of the 28th ACM International Conference on Architectural Support for Programming Languages and Operating Systems, Volume 2},
  pages={574--587},
  year={2023}
}

@inproceedings{kalia2019datacenter,
  title={Datacenter $\{$RPCs$\}$ can be general and fast},
  author={Kalia, Anuj and Kaminsky, Michael and Andersen, David},
  booktitle={16th USENIX Symposium on Networked Systems Design and Implementation (NSDI 19)},
  pages={1--16},
  year={2019}
}

@inproceedings{an2024cxldl,
  title={CXL-DL: A CXL-based Shared Memory Architecture for Deep Learning},
  author={An, Minwoo and Kim, Jinho and Oh, Juhyun and Lee, Myung-Jae and Jung, Myoungsoo},
  booktitle={Proceedings of the 30th IEEE International Symposium on High-Performance Computer Architecture (HPCA)},
  pages={1--14},
  year={2024}
}

@article{kim2023cxlpnm,
  title={CXL-PNM: A CXL-based Processing-near-Memory Accelerator for Large-scale NLP Models},
  author={Kim, Keunsoo and Lee, Sang-Won and Han, Jin-Hee and Park, Gi-Moon and Kim, Young-Cheon and Choi, Jung-Hwan},
  journal={IEEE Computer Architecture Letters},
  volume={22},
  number={2},
  pages={157--160},
  year={2023},
  publisher={IEEE}
}

@article{zhou2024survey,
  title={A survey on efficient inference for large language models},
  author={Zhou, Zixuan and Ning, Xuefei and Hong, Ke and Fu, Tianyu and Xu, Jiaming and Li, Shiyao and Lou, Yuming and Wang, Luning and Yuan, Zhihang and Li, Xiuhong and others},
  journal={arXiv preprint arXiv:2404.14294},
  year={2024}
}

@inproceedings{sheng2023flexgen,
  title={High-throughput Generative Inference of Large Language Models with a Single GPU},
  author={Sheng, Ying and Zheng, Lianmin and Yuan, Binhang and Li, Zhuohan and Maxane, Max and Chen, Beidi and Fu, Ce and Xie, Zhiqiang and Chen, Beidi and Chen, Jiaao and others},
  booktitle={International Conference on Machine Learning (ICML)},
  year={2023}
}

@inproceedings{zhu2024nanoflow,
  title={NanoFlow: Towards Optimal Capacity and Efficiency in Throughput-Oriented {LLM} Serving},
  author={Zhu, Kan and Zhao, Wenyi and others},
  booktitle={arXiv preprint arXiv:2408.12757},
  year={2024}
}

@inproceedings{zhong2024distserve,
  title={DistServe: Disaggregating Prefill and Decoding for Goodput-optimized Large Language Model Serving},
  author={Zhong, Yinmin and Liu, Shengyu and Chen, Junda and Huang, Jian and others},
  booktitle={Proceedings of the 17th USENIX Symposium on Operating Systems Design and Implementation (OSDI)},
  year={2024}
}

@misc{vicuna2023,
    title = {Vicuna: An Open-Source Chatbot Impressing GPT-4 with 90\%* ChatGPT Quality},
    url = {https://lmsys.org/blog/2023-03-30-vicuna/},
    author = {Chiang, Wei-Lin and Li, Zhuohan and Lin, Zi and Sheng, Ying and Wu, Zhanghao and Zhang, Hao and Zheng, Lianmin and Zhuang, Siyuan and Zhuang, Yonghao and Gonzalez, Joseph E. and Stoica, Ion and Duan, Haier},
    month = {March},
    year = {2023}
}

@article{liao2025ub,
  title={Ub-mesh: a hierarchically localized nd-fullmesh datacenter network architecture},
  author={Liao, Heng and Liu, Bingyang and Chen, Xianping and Guo, Zhigang and Cheng, Chuanning and Wang, Jianbing and Chen, Xiangyu and Dong, Peng and Meng, Rui and Liu, Wenjie and others},
  journal={IEEE Micro},
  year={2025},
  publisher={IEEE}
}

@article{arsid2025ultra,
  title={Ultra Ethernet and UALink: Next-Generation Interconnects for AI Infrastructure},
  author={Arsid, Rajesh},
  journal={IJSAT-International Journal on Science and Technology},
  volume={16},
  number={2},
  year={2025},
  publisher={International Research Publication and Journals}
}

\newpage
\appendix








\section{Experimental Setup}
\label{app:exp-details}
This section provides the detailed hardware and software configuration used in our experiments. To ensure reproducibility, we specify the hardware specifications, memory pool setups, and framework parameters.

\subsection{Hardware Platform}
All experiments were conducted on a high-performance server equipped with the hardware specifications summarized in Table \ref{tab:hardware_specs}.

\begin{table}[h]
\centering
\caption{Hardware Specifications}
\label{tab:hardware_specs}
\begin{tabular}{ll}
\toprule
\textbf{Component} & \textbf{Configuration} \\ \midrule
Operating System   & Ubuntu 22.04 (Kernel 6.2.0-1015) \\
CPU                & $2 \times$ Intel(R) Xeon(R) Platinum 8575C \\
L3 Cache           & 640 MiB (320 MiB per CPU) \\
System DRAM        & 2 TB ($32 \times$ DDR5 4800 MT/s 64 GB) \\
GPU                & $8 \times$ NVIDIA H20 (96 GB) \\
Interconnect       & $4 \times$ PCIe Switch (PCIe 5.0) \\ \bottomrule
\end{tabular}
\end{table}

\subsection{Memory Pool Configurations}

\textbf{CXL Memory Pool.} 
The CXL infrastructure is built around an XConn XC50256 CXL switch. We utilize two PCIe 5.0 x16 CXL adapters to interface with the switch. The CXL memory pool totals 2 TB, composed of two CXL Type-3 devices (each featuring $4 \times$ DDR5 1200 MT/s modules for a total of 256 GB per device). Data Direct I/O (DDIO) is disabled to bypass the Last Level Cache (LLC), ensuring cache coherency across nodes and direct memory access.

\textbf{RDMA Memory Pool.} 
For the RDMA baseline, we utilize eight ConnectX-7 single-port 100Gbps NICs. We emulate the RDMA pool by storing the KV cache in local DRAM. Data transfers are routed through the RDMA NICs in a loopback configuration to reflect the latency and throughput characteristics of a network-attached memory pool.
As physical cross-node RDMA typically incurs higher latency due to additional network hops and cable propagation, our loopback configuration provides an idealized, best-case baseline for RDMA. Consequently, the performance gains reported for SAC represent a conservative estimate of its real-world advantages.

\subsection{Software and Framework Configuration}

\textbf{SGLang Settings.} 
We deploy SGLang with the following hyperparameters:
\begin{itemize}
    \item Tensor Parallelism (TP): 8
    \item Data Parallelism (DP Attention): 8
    \item Memory Configuration: \texttt{bf16} precision, \texttt{mem-static-fraction} set to 0.85.
\end{itemize}

\textbf{HiSparse Settings.} 
We configure HiSparse with a \textit{top-k} value of 2048 and a device buffer pool of 6144. As standard HiSparse currently does not support Radix Cache, we implemented a custom Radix Cache integration within HiSparse. In this configuration, the KV cache is entirely offloaded to the designated memory backend (CXL or RDMA).

\newpage
\subsection{Third-Party Assets and Licenses}
\label{app:thirdparty-licenses}
\noindent Table~\ref{tab:thirdparty-licenses} summarizes software, models, and data used in evaluation (see also Section~\ref{sec:eval}). We use each asset only under its stated terms.

\begin{table}[h]
\centering
\caption{Third-party assets and their licenses (informative).}
\label{tab:thirdparty-licenses}
\begin{tabular}{llp{5.9cm}}
\toprule
\textbf{Asset} & \textbf{License (as referenced)} & \textbf{Notes} \\ \midrule
SGLang + HiSparse path~\cite{zheng2024sglang,zheng2026hisparse} & Apache License 2.0 & SGLang's official repository ships under Apache-2.0; HiSparse sparse-serving behavior is used within this same software stack as described in~\cite{zheng2026hisparse}. \\
DeepSeek-V3.2 weights~\cite{deepseekv32} & MIT License & As stated on the Hugging Face model card for \texttt{deepseek-ai/DeepSeek-V3.2}. \\
ShareGPT-format traces~\cite{vicuna2023} & Dataset-dependent & Requests follow the ShareGPT-style setting discussed with Vicuna~\cite{vicuna2023}; we adhere to the license of the specific ShareGPT-format snapshot used when sampling. \\
\bottomrule
\end{tabular}
\end{table}

\newpage

\section{Extended Literature Review}

\subsection{Related Works: CXL-based KV Cache Management System}

The limitations of RDMA-based disaggregated memory pools, specifically high access latency, complex communication protocols, and synchronization overhead, have recently spurred the development of CXL-based architectures for LLM serving. Two prominent systems, \textbf{Beluga}~\cite{yang2025beluga} and \textbf{TraCT}~\cite{yoon2025tract}, have established the foundation for using CXL as a high-performance substrate for KV cache.

\subsubsection{Beluga: Scalable Shared Memory via CXL Switches}
Beluga~\cite{yang2025beluga} is the first system to leverage CXL 2.0 switches to enable GPU clusters to access a shared, disaggregated memory pool. By replacing the network-based RDMA protocol with a memory-semantic interface, Beluga allows GPUs to perform direct peer-to-peer (P2P) transfers and fine-grained, non-contiguous access via custom CUDA kernels. 

A key contribution of Beluga is its characterization of the ``indirect host-staged data path'' in RDMA, where data must move through a host ``bounce buffer,'' incurring significant latency. Beluga eliminates these extra copies and simplifies the control path by unifying data transfer within the GPU's native CUDA stream. While Beluga acknowledges the efficiency of CXL for sparse KV cache access patterns, its primary focus remains the management of large-scale, contiguous KV cache blocks within a unified address space to simplify scheduling and capacity scaling.

\subsubsection{TraCT: Rack-Scale Serving with Direct GPU-CXL DMA}
TraCT~\cite{yoon2025tract} focuses on disaggregated LLM serving, specifically separating the compute-intensive prefill phase from the latency-critical decode phase. It uses CXL shared memory as both a KV-transfer substrate and a rack-wide prefix-aware cache. Unlike previous systems that rely on RDMA to move KV tensors between prefill and decode workers, TraCT enables GPUs to write and read KV blocks directly via CXL load/store and DMA operations, entirely eliminating the NIC hop. 

TraCT provides critical software solutions for operating on non-coherent CXL Type-3 devices through three key mechanisms. First, it implements two-tier synchronization to bound contention by utilizing local DRAM locks in conjunction with a global shared-memory lock array. To ensure data visibility, TraCT employs fine-grained \texttt{clflush} instructions to guarantee metadata consistency across nodes . Finally, it facilitates zero-copy transfers by pinning the CXL memory region, which allows the CUDA runtime to treat and access it directly as page-locked host memory.

\subsubsection{Distinguishing SAC from Existing CXL Systems}
While Beluga and TraCT establish CXL as a superior alternative to RDMA for bulk KV block transfers, they primarily optimize for Dense Attention models where KV data is treated as monolithic blocks to be published or consumed. Our work, SAC (Sparse Attention on CXL), introduces a fundamental shift:

\begin{enumerate}
    \item \textbf{From Transfer to On-Demand Fetching}: Whereas TraCT and Beluga focus on moving KV blocks, SAC utilizes CXL's near-DRAM latency to fetch only the \textbf{top-$k$ KV entries} in real-time during the attention calculation. This addresses the Transmission Bottleneck (P1) that remains even in fast bulk-transfer systems.
    \item \textbf{Mandatory vs. Optional Substrate}: In dense models, CXL is a performance optimizer; for Sparse Attention, SAC demonstrates that CXL is a necessity. By residing the full KV cache in the CXL pool and fetching only active entries, SAC resolves the Local Memory Wasting (P2) inherent in the ``full prefetching'' strategy used by RDMA-based systems.
    \item \textbf{Fine-Grained Integration}: Unlike the block-level management in TraCT, SAC integrates directly with the SGLang runtime to handle the dynamic, runtime-determined indices of sparse attention models, utilizing CXL's zero-protocol overhead for discrete, cache-line granularity loads.
\end{enumerate}

In essence, while prior work focuses on CXL as a \textbf{faster storage tier}, SAC treats CXL as a \textbf{direct extension of the GPU's memory hierarchy} specifically tailored for the non-contiguous access patterns of modern sparse LLM architectures.

\subsection{CXL Techniques and its Applications}
\label{app:cxl_tech_apps}

\textbf{Protocol Evolution and Standards.} Compute Express Link (CXL) was established by the CXL Consortium in 2019 to overcome the memory wall by leveraging the PCIe physical layer. While CXL 1.1 primarily supported point-to-point host-to-device connections, the protocol has significantly matured with subsequent versions. CXL 2.0 introduced hardware-managed switching and memory pooling capabilities, while CXL 3.0/3.1 further expanded these into a fabric-based architecture. This evolution allows CXL to provide high-bandwidth, low-latency memory sharing with hardware-managed load/store semantics, enabling access at cache-line granularity which is uniquely suited for fine-grained, sparse memory lookups.

\textbf{Applications in Database Systems.} In the domain of database systems, CXL is increasingly used as a compelling alternative to RDMA for cross-node memory access. Tigon ~\cite{cxl_db_2} introduces a distributed database architecture specifically optimized for CXL pods, achieving high performance through the protocol's native memory semantics. Similarly, Yang et al.~\cite{cxl_db_1} explore unlocking the potential of CXL for disaggregated memory in cloud-native databases. These systems benefit from CXL to simplify data paths and transaction management compared to traditional network-based approaches.

\textbf{Adoption in Cloud Platforms.} Cloud infrastructure providers have adopted CXL to mitigate memory stranding and enhance resource utilization. Pond~\cite{cxl_cloud_3} demonstrated an early production-scale CXL-based memory pooling system. Memtunnel~\cite{cxl_cloud_2} shifted the research focus toward the broader challenge of rack-scale orchestration. Zhong et al.~\cite{cxl_cloud_1} focuses on managing memory tiers within virtualized environments. RCMP~\cite{cxl_db_3} shows that CXL-based memory disaggregation significantly reduces software overhead and improves CPU efficiency in cloud-scale deployments.

\textbf{Application to LLM Serving.} Beyond general cloud workloads, CXL is becoming a critical enabler for Large Language Model (LLM) serving and Deep Learning (DL) acceleration. While Beluga \cite{yang2025beluga} and TraCT \cite{yoon2025tract} primarily focus on prefix KV cache management for dense models, other works explore broader architectural optimizations. For instance, CXL-DL~\cite{an2024cxldl} investigates using CXL-attached memory to expand capacity for massive embedding tables in recommendation systems, which share the characteristic of sparse memory access with certain LLM configurations. Additionally, hardware-software co-designs like CXL-PNM~\cite{kim2023cxlpnm} explore processing-near-memory within CXL Type-3 devices to accelerate LLM inference by reducing data movement. 

\subsection{Hardware Cost Analysis: CXL vs. RDMA}

To quantitatively evaluate the cost-effectiveness of the SAC architecture, we provide a comparison between RDMA-based networking and CXL-based memory pooling, based on commercial hardware specifications. As shown in Table \ref{table:cost}, CXL-based components significantly reduce the cost of host interface cards and switches at the device level.

\begin{table}[h]
\centering
\small 
\caption{Hardware Cost Analysis~\cite{yang2025beluga}}
\label{table:cost}
\begin{tabular}{lll} 
\hline
\textbf{Component} & \textbf{RDMA-based (RoCE)} & \textbf{CXL-based} \\ \hline
Interface on Host & ConnectX-7 & PCIe/CXL Adapter \\ 
PCIe Lanes & x16 & x16 \\ 
Price & \$1,745 & \$210 \\ \hline
Switch & Mellanox RoCE Switch & XConn CXL Switch \\ 
Ports/Capacity & 40 x 200Gbps & 32 x PCIe 5.0 x16 \\ 
Price & \$16,000 & \$5,800 \\ \hline
\textbf{\$/(64GB/s)} & \textbf{\$800} & \textbf{\$218.75} \\ \hline
\end{tabular}
\end{table}

Beyond direct component pricing, SAC offers significant system-level cost benefits:
\begin{itemize}
    \item \textbf{Memory Density and Type}: Unlike traditional DIMM memory in the host, CXL memory devices can use lower-density chips, reducing the cost per GB.
    \item \textbf{Resource Disaggregation}: CXL memory pools separate memory from CPU resources, allowing multiple servers to share memory and improve utilization in cloud environments.
    \item \textbf{Local Memory Reduction}: By keeping the full KV cache in the CXL pool and fetching only the required top-$k$ entries on demand, SAC eliminates the need for over-provisioning expensive TB-level local DRAM on every compute node, further lowering the total cost of ownership (TCO) for sparse attention model serving.
\end{itemize}
\newpage

\section{SGLang HiSparse: The Implementation Basis of SAC}
\label{app:hisparse}

The performance of SAC (Sparse Attention on CXL) is fundamentally built upon the architectural principles of \textbf{HiSparse}~\cite{zheng2026hisparse}, a hierarchical memory management mechanism developed for the SGLang~\cite{zheng2024sglang} framework. This section details the motivation and implementation of HiSparse.

\subsection{Motivation: Breaking the Memory Capacity Bottleneck}
While sparse attention models reduce computational complexity by attending to a subset of tokens, they traditionally do not solve the local memory capacity limitation problem. In standard serving stacks, the KV cache for the full context must remain resident in GPU HBM to ensure low-latency access, even though only a tiny fraction of entries are active during any given decoding step.

\begin{figure}[h] 
  \centering
  \includegraphics[width=0.8\linewidth]{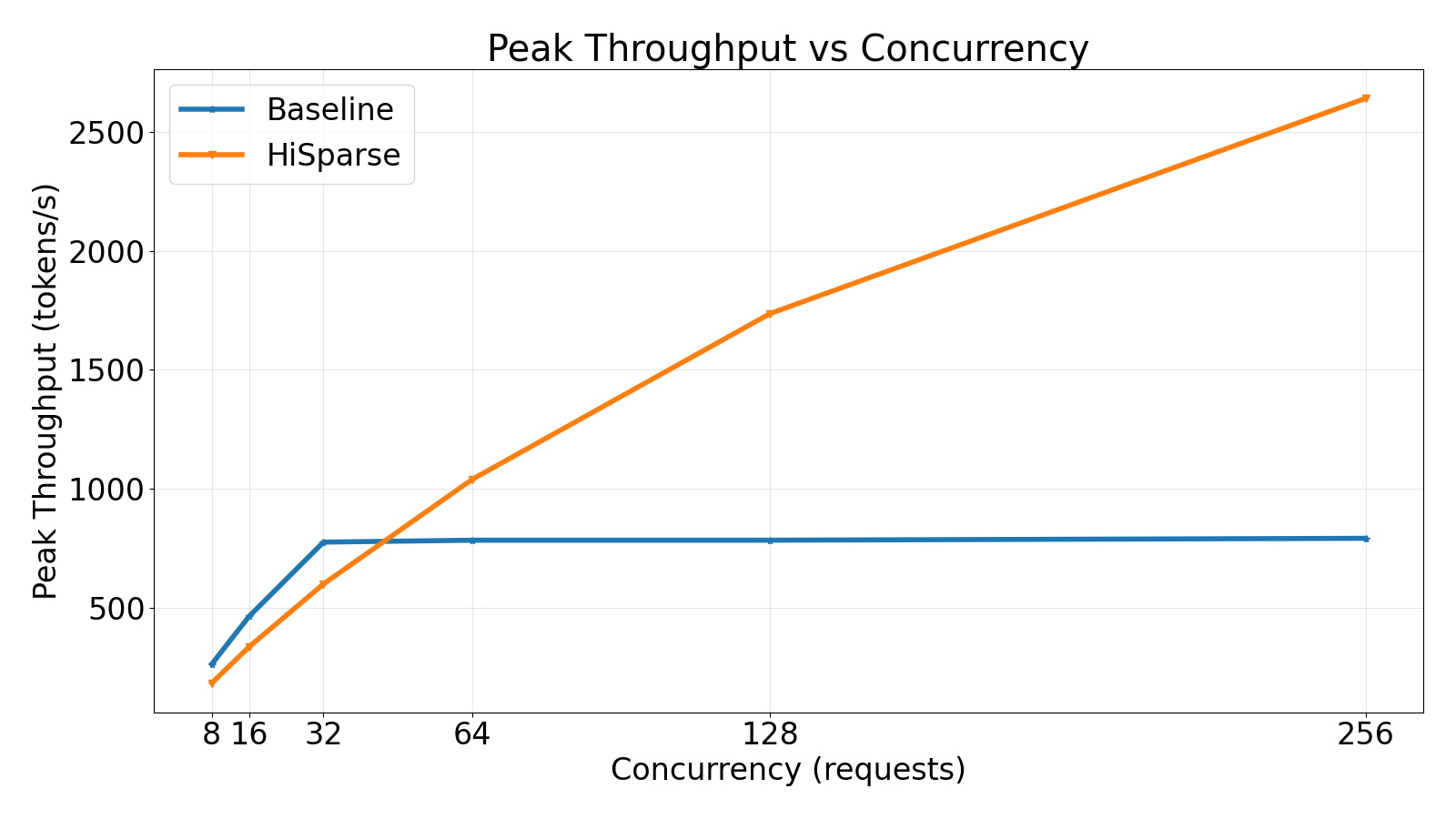}
  \caption{Benchmark results for the GLM-5.1-FP8 model using 32k-input, 8k-output queries on a PD-colocated 8×H200 deployment~\cite{zheng2026hisparse}.}
  \label{fig:hisparse_throughput}
\end{figure}

As illustrated in Figure \ref{fig:hisparse_throughput}, standard sparse attention implementations hit a memory capacity wall early, causing throughput to plateau as concurrency increases. HiSparse addresses this by introducing a hierarchical memory approach that offloads inactive KV cache entries to a secondary memory tier. This allows for near-linear throughput scaling; at 256 concurrent requests, HiSparse achieves over 3$\times$ the throughput of the baseline by decoupling token generation from local HBM capacity.

\subsection{HiSparse Architecture and Workflow}
\begin{figure}[h] 
  \centering
  \includegraphics[width=1.0\linewidth]{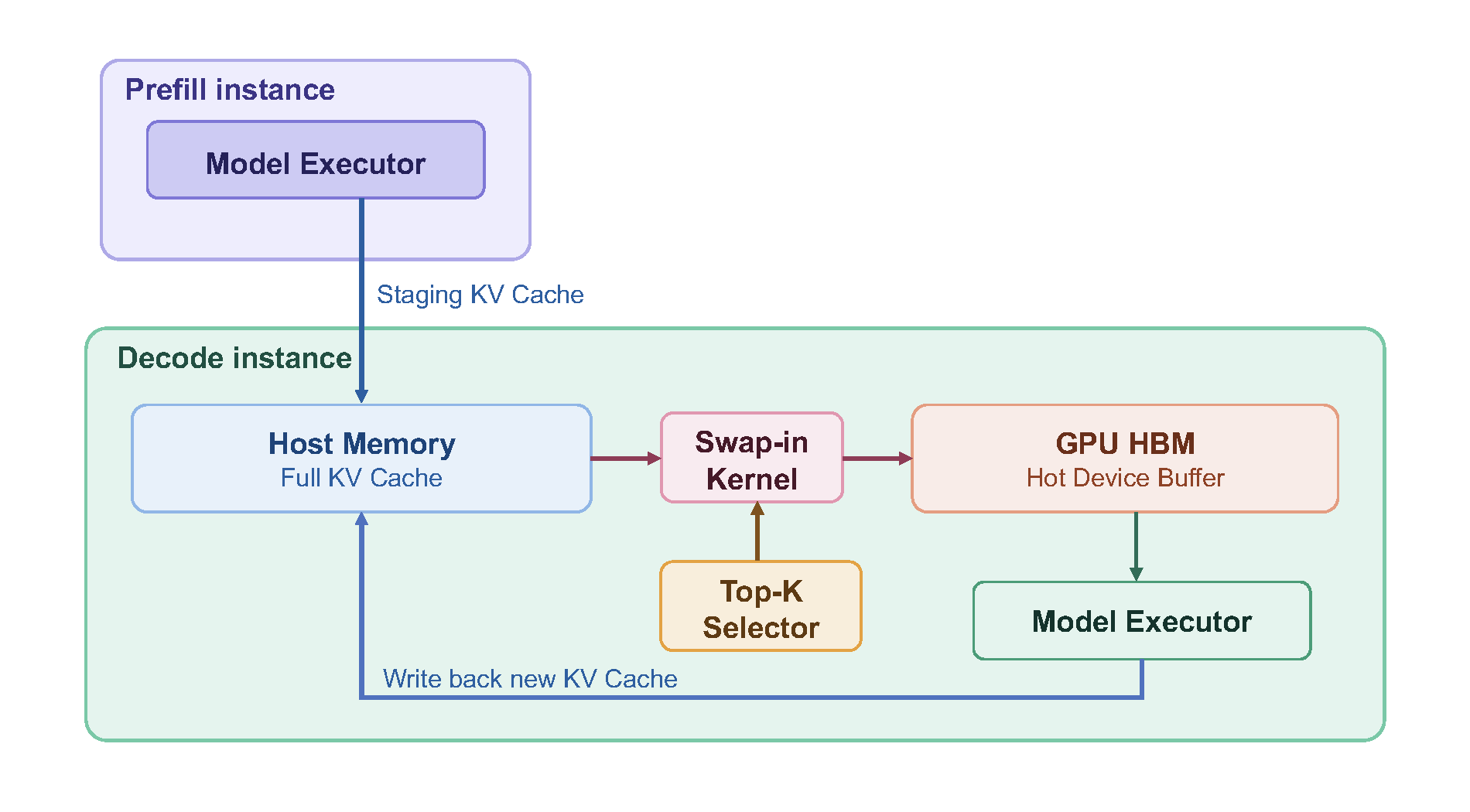}
  \caption{HiSparse workflow~\cite{zheng2026hisparse}.}
  \label{fig:hisparse_workflow}
\end{figure}

The architecture of HiSparse revolves around a proactive offloading and reactive swap-in mechanism. As shown in Figure~\ref{fig:hisparse_workflow}, the system maintains a subset of the KV cache in the local device buffer to minimize data movement on the critical path.

Central to HiSparse is a specialized Swap-in CUDA kernel designed to manage the data flow between the local HBM and the remote memory pool. The kernel performs three critical operations during each decoding step:
\begin{enumerate}
    \item \textbf{Miss Identification}: It identifies which of the dynamically determined Top-$k$ KV entries are missing from the local device buffer.
    \item \textbf{LRU Eviction}: It selects eviction candidates from the hot device buffer using a Least Recently Used (LRU) policy to make space for incoming data.
    \item \textbf{Page Table Update and Fetch}: It updates the internal page table mapping and triggers the fetch of the required KV entries from the remote pool to the local HBM. 
\end{enumerate}

\subsection{SAC Implementation on HiSparse}

SAC orchestrates a hardware-software co-design by coupling the HiSparse framework with a CXL-centric memory hierarchy. This integration enables sparse decoding to leverage a fabric-attached, high-capacity KV tier, effectively decoupling the logical context length from the physical constraints of local memory.

The implementation is characterized by the following architectural components:

\begin{itemize}
    \item \textbf{Hierarchical Memory Abstraction}: SAC redefines the HiSparse KV management stack by establishing CXL-backed disaggregated pools as the capacity tier. This architecture allows the authoritative KV corpus to scale independently of GPU HBM, providing the necessary volume for long-context windows while maintaining a logically unified memory fabric.
    \item \textbf{Unified Access Semantics}: By leveraging the memory-semantic nature of the CXL physical layer, SAC optimizes the CUDA kernels in HiSparse to treat disaggregated memory as a direct extension of its local hierarchy. This bypasses traditional staging overheads and allows for fine-grained, zero-copy interactions between the GPU and the remote KV buffer.
    \item \textbf{Sparsity-Aware KV Fetching}: SAC aligns the data movement strategy with HiSparse. The system performs reactive, sub-block fetching driven by layer-wise top-$k$ selection, which exploits CXL’s low-latency characteristics to satisfy real-time inference requirements with minimal I/O amplification.
\end{itemize}

\newpage

\section{More Experimental Results and Analysis}
\subsection{Statistic Significance Evaluation}
For the main evaluation in Section~\ref{sec:e2e}, we repeat each experiment 3 times and report the average of the results. 
Because the inference process is largely deterministic and conducted in a controlled environment, the experimental results exhibit minimal divergence.
To quantify this stability, we calculate the Coefficient of Variation (CV), defined as $CV = (\sigma / \mu) \times 100\%$, where $\sigma$ is the standard deviation and $\mu$ is the mean. 
Across all test cases (including various context lengths and concurrency levels), the CV for throughput remains below 2.1\%, and the CV for latency metrics (TTFT and TBT) is consistently within 3.5\%.
The small standard deviation relative to the performance gains demonstrates that the observed advantages of SAC are statistically significant and not the result of transient hardware fluctuations or measurement noise.

\subsection{Performance under Varying Output Lengths}
\label{app:output_len}
In the main evaluation in Section~\ref{sec:e2e}, we fixed the output length at 1K tokens. This configuration represents a common production serving scenario where the generated output is significantly shorter than the input context. As AI agent workloads become increasingly mainstream, this disparity grows further, with input-to-output ratios often reaching 100:1 or even 200:1~\cite{mooncake}. In these "long-input, short-output" scenarios, the inefficiencies of RDMA-based disaggregated systems—the transmission bottleneck and local memory wastage—become even more pronounced, as the entire KV cache must be fetched locally only to generate a relatively small number of tokens.

\begin{figure}[H]
\centering
\includegraphics[width=0.99\linewidth]{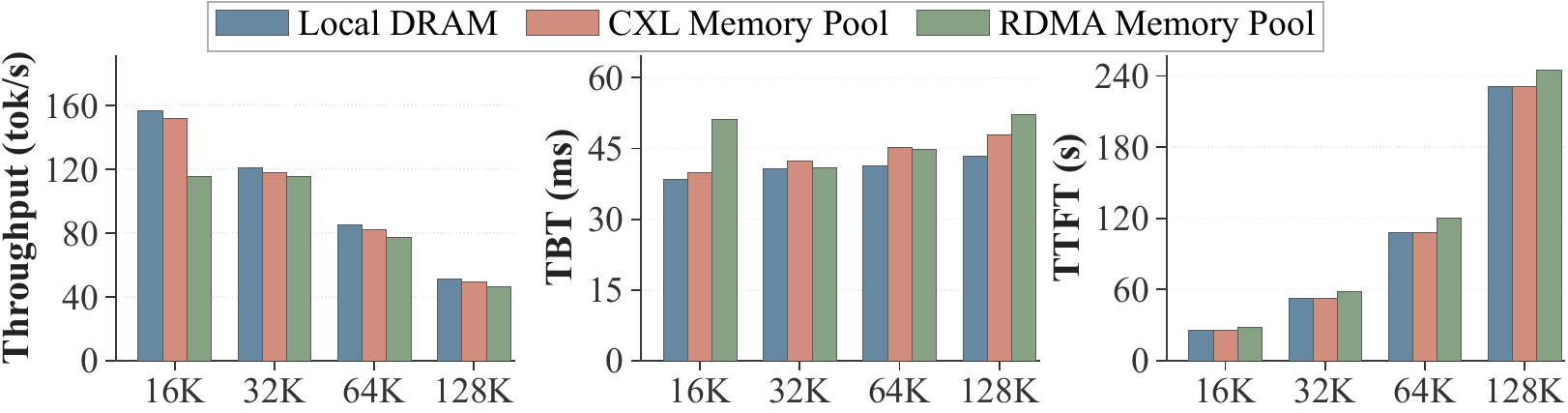}
\vspace{-0.5em}
\caption{Performance comparison with 2K output length: Round-1 (Prefill).}
\label{fig:exp_2k}
\end{figure}

\begin{figure}[H]
\centering
\includegraphics[width=0.99\linewidth]{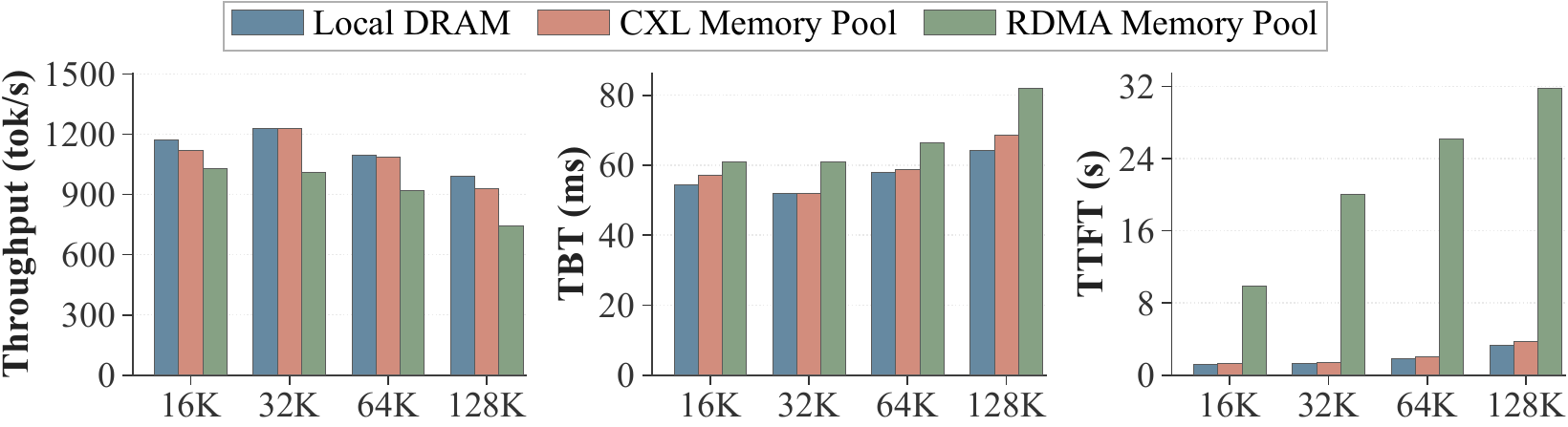}
\vspace{-0.5em}
\caption{Performance comparison with 8K output length: Round-2 (Decoding).}
\vspace{-10pt}
\label{fig:exp_8k}
\end{figure}

To evaluate the robustness of SAC across diverse workloads, we measure performance across various output lengths including 2K, 4K, and 8K tokens. The results are illustrated in Figures~\ref{fig:exp_2k} through \ref{fig:exp_8k}.

\begin{figure}[H]
\centering
\includegraphics[width=0.99\linewidth]{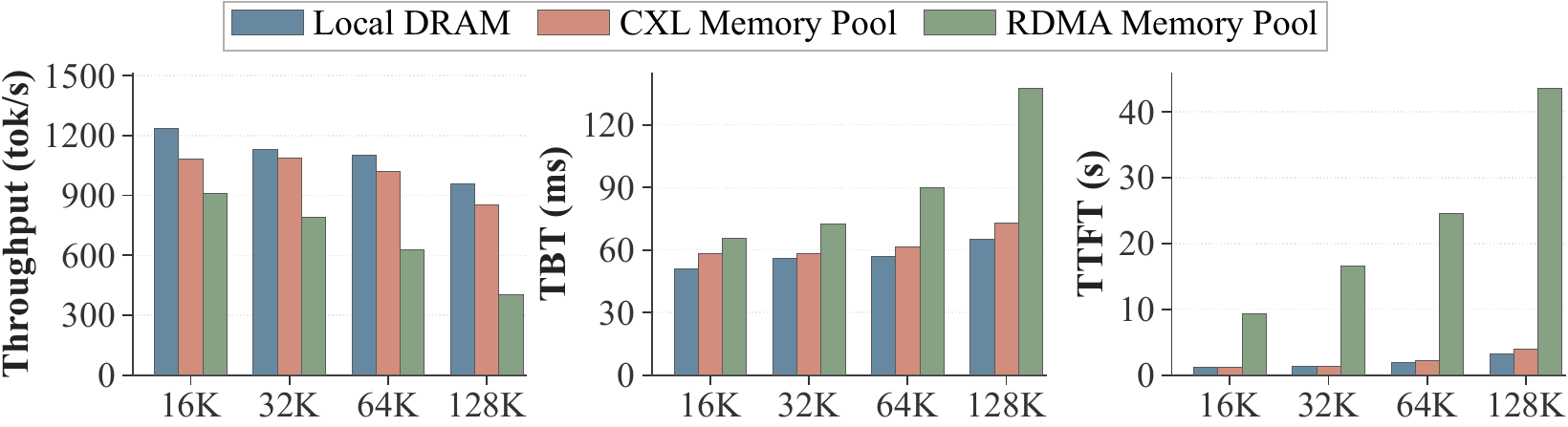}
\vspace{-0.5em}
\caption{Performance comparison with 2K output length: Round-2 (Decoding).}
\end{figure}

\begin{figure}[H]
\centering
\includegraphics[width=0.99\linewidth]{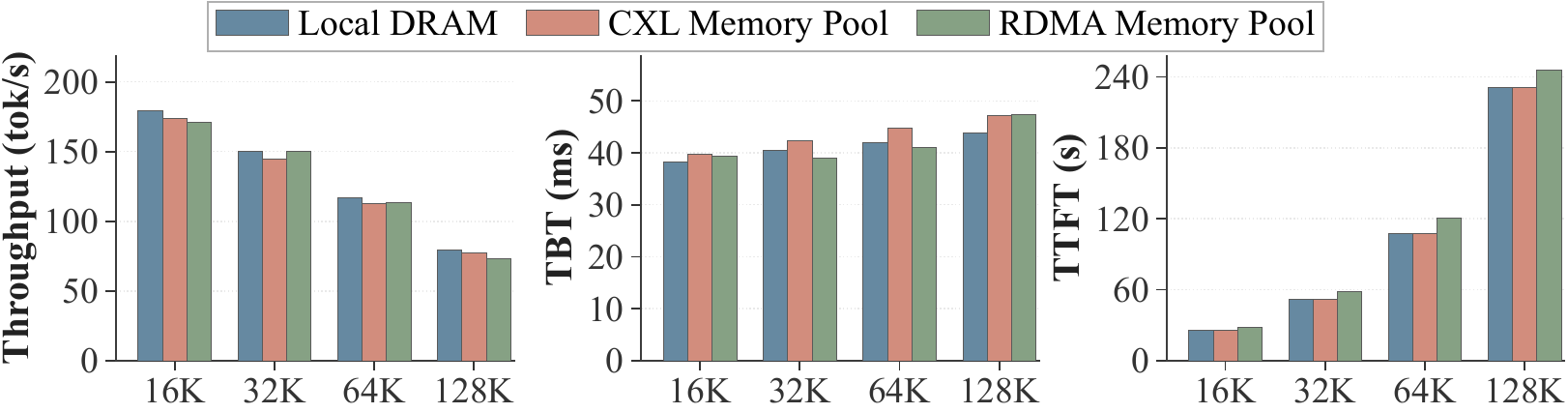}
\vspace{-0.5em}
\caption{Performance comparison with 4K output length: Round-1 (Prefill).}
\vspace{-10pt}
\end{figure}

\begin{figure}[H]
\centering
\includegraphics[width=0.99\linewidth]{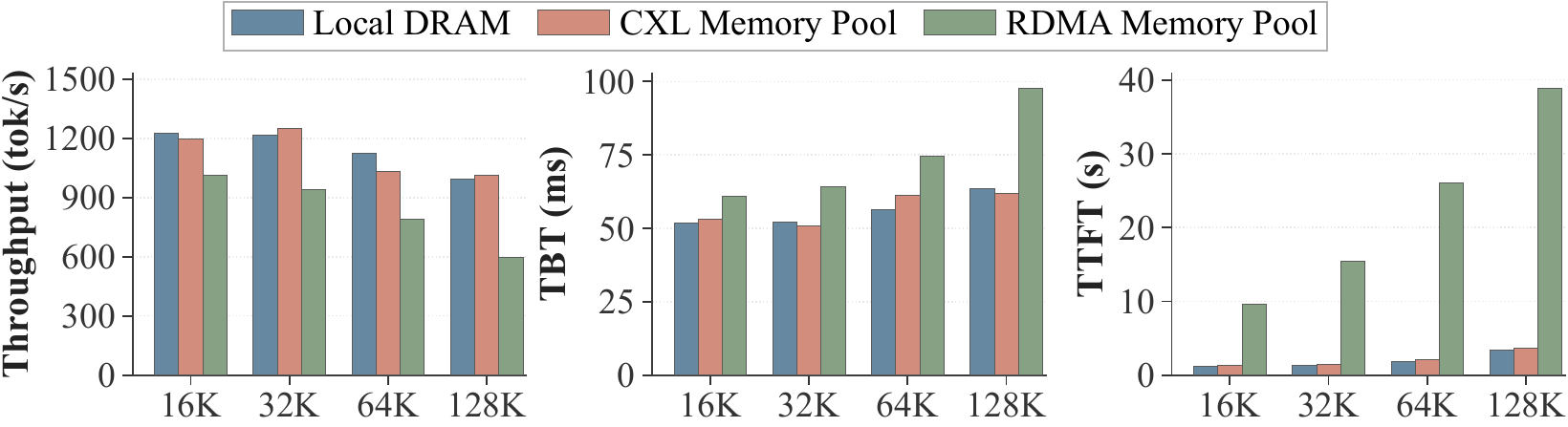}
\vspace{-0.5em}
\caption{Performance comparison with 4K output length: Round-2 (Decoding).}
\vspace{-10pt}
\end{figure}

\begin{figure}[H]
\centering
\includegraphics[width=0.99\linewidth]{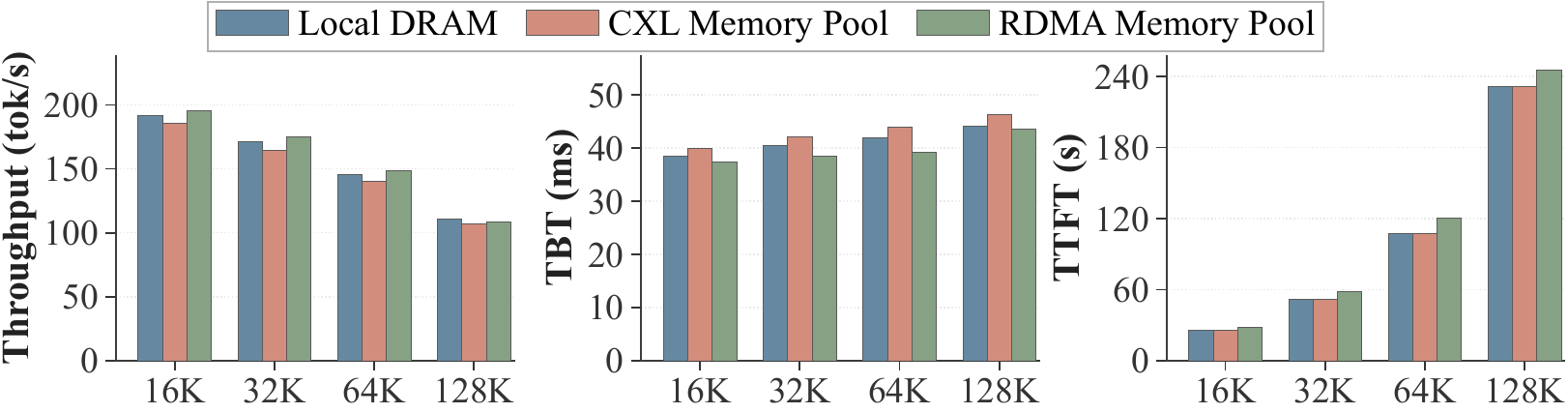}
\vspace{-0.5em}
\caption{Performance comparison with 8K output length: Round-1 (Prefill).}
\vspace{-10pt}
\end{figure}

The experimental results consistently align with the findings in Section~\ref{sec:e2e}. In Round-1 (prefill stage), all backends exhibit nearly identical performance, as the stage remains primarily compute-bound on the GPU. In Round-2 (decoding stage), SAC maintains a clear throughput advantage over RDMA across all tested output lengths, including the 8K configuration.

Notably, we observe that the throughput advantage of SAC is most significant at shorter output lengths. This is because for shorter generations, the massive fixed overhead of fetching the entire prefix KV cache via RDMA dominates the total request latency. As output length increases, this initial "transmission tax" is amortized over a larger number of generated tokens; however, SAC continues to outperform RDMA due to its superior handling of PCIe bus contention and lower TBT, as discussed in the main evaluation.

\subsection{Tail Latency Analysis}
\label{app:tail}

\begin{figure}[H]
\centering
\includegraphics[width=0.99\linewidth]{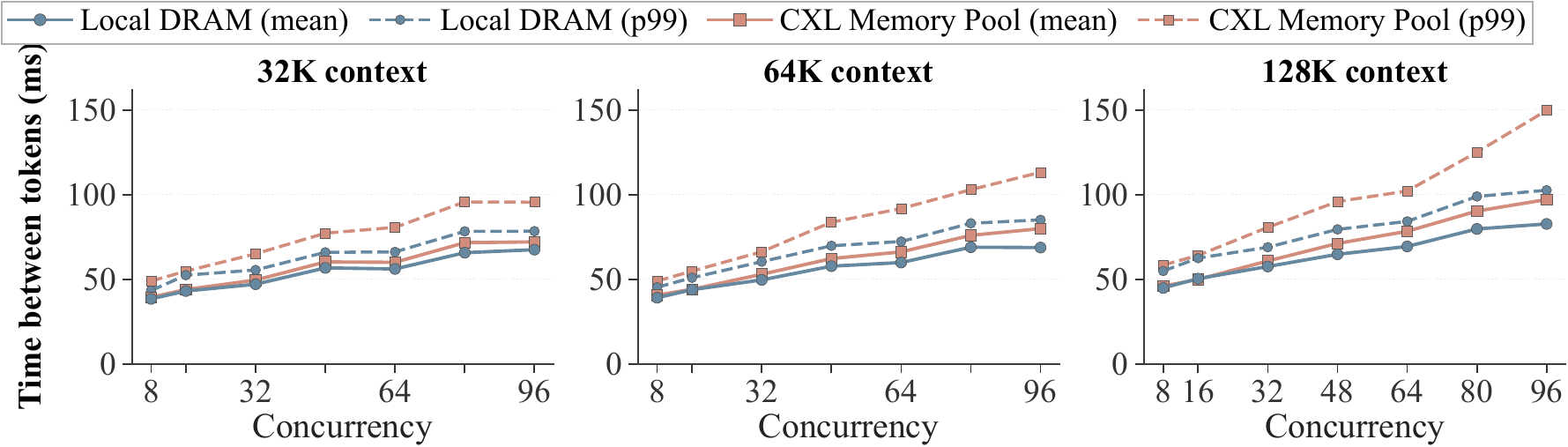}
\vspace{-0.5em}
\caption{TBT Tail Latency: Comparison of Mean and p99 TBT across increasing concurrency.}
\label{fig:appendix-tail-tbt}
\end{figure}

\begin{figure}[H]
\centering
\includegraphics[width=0.99\linewidth]{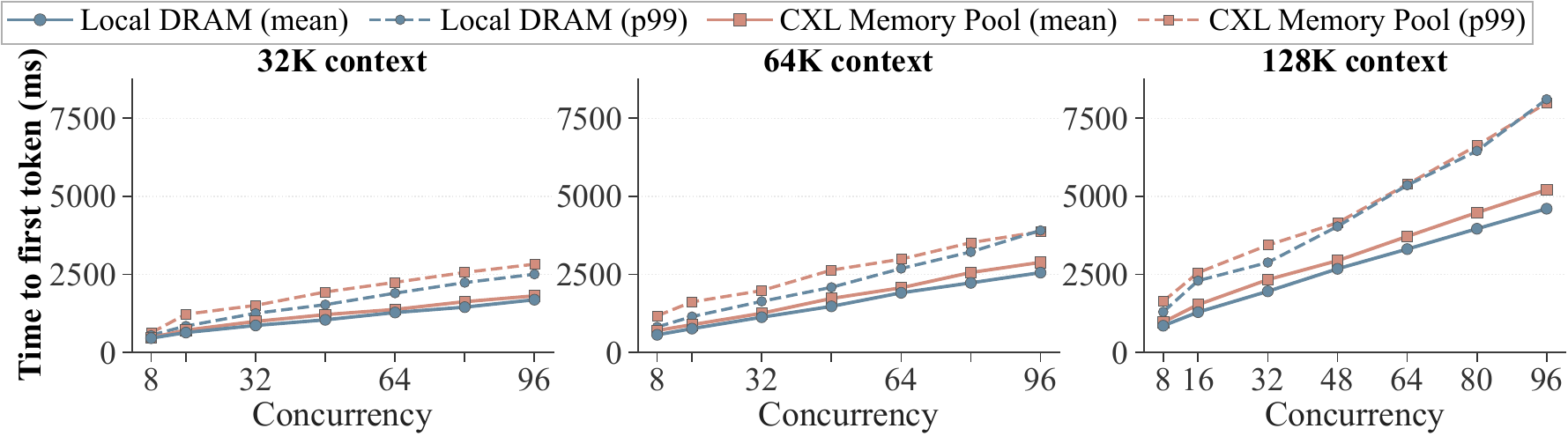}
\vspace{-0.5em}
\caption{TTFT Tail Latency: Comparison of Mean and p99 TTFT across increasing concurrency.}
\label{fig:appendix-tail-ttft}
\end{figure}

While the main evaluation focuses on average throughput and mean latency, this section quantifies the tail behavior of SAC under increasing concurrency. Figures~\ref{fig:appendix-tail-tbt} and \ref{fig:appendix-tail-ttft} illustrate the distribution of time between tokens (TBT) and time to first token (TTFT), respectively.

We observe that while the mean TBT grows modestly across all backends as concurrency increases, the p99 metrics for both TBT and TTFT rise at a significantly higher rate. This divergence indicates intensified request-level contention and latency variance under heavy loads. Notably, the CXL-based memory pool exhibits a wider gap between mean and p99 latencies compared to the local DRAM baseline at equivalent concurrency levels. This behavior is consistent with memory-side contention and arbitration overhead within the CXL fabric, which scales non-linearly with increasing request concurrency. 

\subsection{Request-level Throughput}
\label{app:req-throughput}

\begin{figure}[H]
\centering
\includegraphics[width=0.7\linewidth]{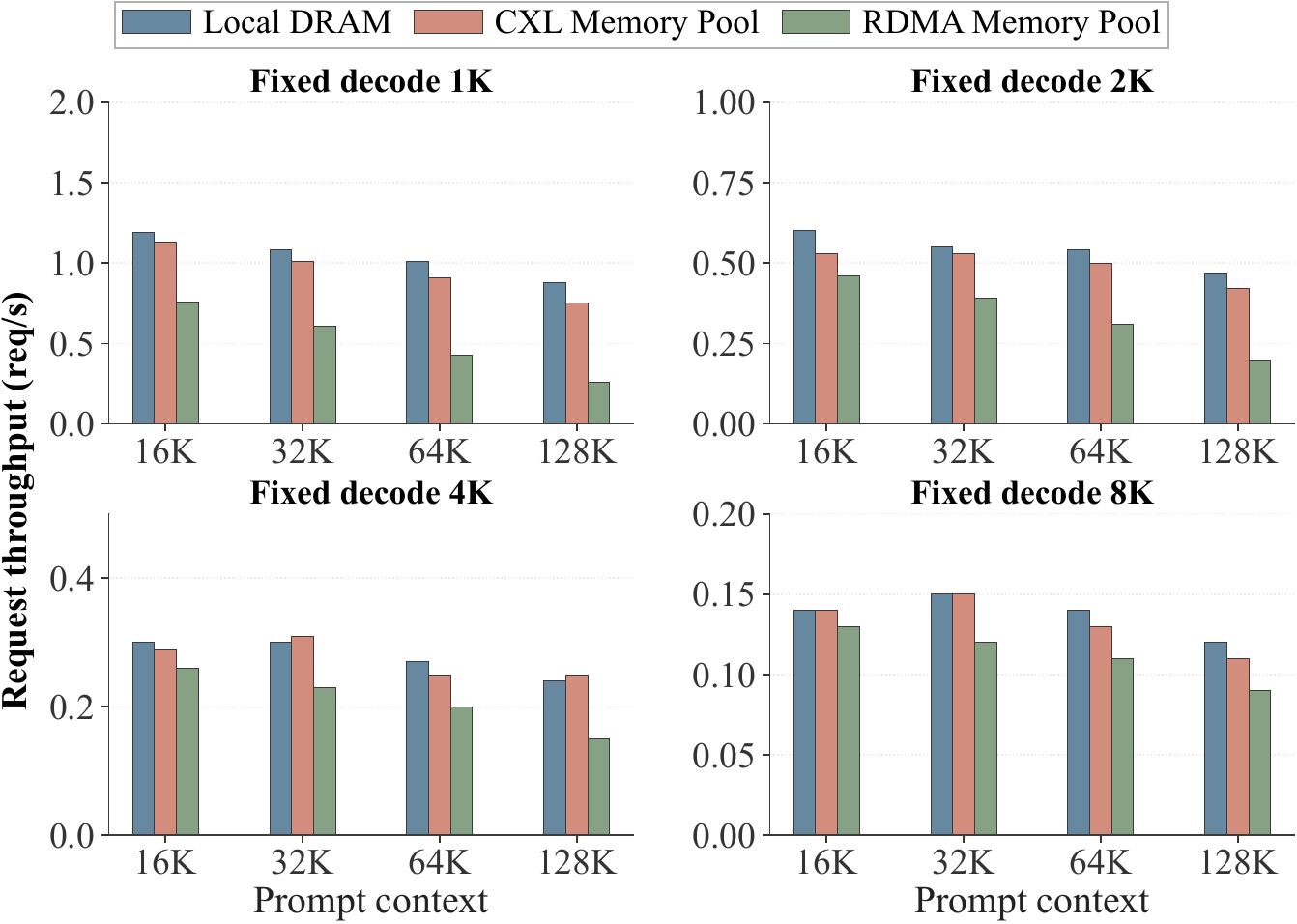}
\vspace{-0.5em}
\caption{Request throughput (req/s) comparison across different backends and output lengths.}
\label{fig:app-req-throughput}
\end{figure}

While the primary evaluation focuses on token-level throughput (tokens/s), we provide the corresponding request throughput (requests/s) here to offer a more comprehensive performance profile. As illustrated in Figure~\ref{fig:app-req-throughput}, SAC achieves request throughput that closely approaches the local DRAM baseline across all evaluated output lengths. Furthermore, SAC significantly outperforms the RDMA-based disaggregated pool. This consistency between token-level and request-level metrics confirms the efficiency gains provided by SAC.

\end{document}